\documentclass{article}
\usepackage{amsmath}
 \usepackage{rotating}
\usepackage{arxiv}
\usepackage{ragged2e}
\usepackage{float} 
\usepackage[table]{xcolor} 
\usepackage{dsfont} 
\usepackage[utf8]{inputenc} 
\usepackage[T1]{fontenc}    
\usepackage{hyperref}       
\usepackage{url}            
\usepackage{booktabs}       
\usepackage{amsfonts}       
\usepackage{nicefrac}       
\usepackage{microtype}      
\usepackage{lipsum}
\usepackage{xcolor}
\usepackage{graphicx}
\graphicspath{ {./images/} }

\title{Evaluation of Performance Measures for Qualifying Flood Models with Satellite Observations}

\author{
 Jean-Paul Travert \\
  EDF R\&D, Laboratoire National d’Hydraulique et Environnement (LNHE), Chatou, France\\
  Laboratoire d’Hydraulique Saint-Venant (LHSV), Ecole des Ponts, EDF R\&D,
Chatou, France\\
  \texttt{jean-paul.travert@edf.fr} \\
   \And
 Sébastien Boyaval \\
  Laboratoire d’Hydraulique Saint-Venant (LHSV), Ecole des Ponts, EDF R\&D,
Chatou, France\\
  MATHERIALS, Inria, Paris, France\\
  \And
 Cédric Goeury\\
  EDF R\&D, Laboratoire National d’Hydraulique et Environnement (LNHE), Chatou, France\\
  Laboratoire d’Hydraulique Saint-Venant (LHSV), Ecole des Ponts, EDF R\&D,
Chatou, France\\
  \And
 Vito Bacchi\\
  EDF R\&D, Laboratoire National d’Hydraulique et Environnement (LNHE), Chatou, France\\
\And
   Fabrice Zaoui\\
  EDF R\&D, Laboratoire National d’Hydraulique et Environnement (LNHE), Chatou, France\\
}

\begin{document}
\begin{justify}
\maketitle
\begin{abstract}
This work discusses how to choose \emph{performance measures} to compare numerical simulations of a flood event with one satellite image, e.g., in a model calibration or validation procedure. A series of criterion are proposed to evaluate the sensitivity of performance measures with respect to the flood extent, satellite characteristics (position, orientation), and measurements/processing errors (satellite raw values or extraction of the flood maps). Their relevance is discussed numerically in the case of one flooding event (on the Garonne River in France in February 2021), using a distribution of water depths simulated from a shallow-water model parameterized by an uncertain friction field. After identifying the performance measures respecting the most criteria, a correlation analysis is carried out to identify how various performance measures are similar. Then, a methodology is proposed to rank performance measures and select the most robust to observation errors. The methodology is shown useful at identifying four performance measures out of 28 in the study case. Note that the various top-ranked performance measures do not lead to the same calibration result as regards the friction field of the shallow-water model. The methodology can be applied to the comparison of any flood model with any flood event.
\end{abstract}


\section{Introduction}
To assess in real-time the state of a current inundation, or to predict likely future floods, motions of river waters are often simulated by means of numerical models. But such numerical models of our environment only approximate the real-world physical processes. They constantly need being corrected, i.e. calibrated to observations, in order to maintain their forecasting capacity, and the calibration of numerical models with observations is currently challenged by the increase in IT resources. As regards flood models for instance, it is now becoming standard to compare them with satellite images \cite{horritt2000calibration,wood2016calibration,dasgupta2021mutual}, while flood simulations have been calibrated on pointwise gauges during the past fifty-years. The increasing availability of observations of new type (spatially distributed, as images) is the reason why we need new comparison tools between simulations and observations.
This is all the more so as it has been demonstrated that the choice of mathematical tools to evaluate simulations with respect to observations strongly impacts the predictions \cite{beven2006manifesto}. 

This work discusses how to choose \emph{performance measures} to compare numerical simulations of a flood event with one satellite image. 

Performance measures (also called calibration metrics, scores, verification metrics, distance, dissimilarity measures, etc.) are scalar functions that quantify differences between numerical values of a mathematical object, e.g., simulated and observed rasters (grid of pixels). Murphy \cite{murphy1993good} distinguishes three considerations to clarify the goodness of a simulation: the match between the judgements of practitioners and the simulations, the statistical match between the simulations and the observations, and the economic/environmental priorities of the decision makers. This paper is focused on the second issue concerning the quality quantification between numerical simulations and observations based on performance measures.

The choice of an appropriate performance measure depends on the nature of the observations and/or the characteristics of the model. At present, the main data for calibrating flood models in an operational context is water depths, which are measured along the channel at in-situ gauge stations. However, this is few data for the calibration of Two-Dimensional (2D) flood models. Sometimes, measured water marks are available in the floodplains (e.g., Vigicrues collects water marks during floods in France) \cite{aronica1998uncertainty,ballesteros2011calibration}, or satellite altimetric data  \cite{domeneghetti2014use}, still at low spatial resolution however. In the last two decades, with the increasing number of remote sensing observations (aerial photographs or satellite observations), new high-resolution data is available during floods.  This data is leveraged to quantify the quality of model simulations especially in the floodplains \cite{horritt2000calibration,wood2016calibration,horritt2002evaluation,di2009technique,hall2011bayesian,neal2012subgrid}. Among these data, Synthetic Aperture Radar (SAR) images are widely used for model quality assessment. Indeed, with satellite missions such as Sentinel-1, free data is available and SAR images allow detecting water even during cloudy conditions and at night \cite{tarpanelli2022effectiveness}. From these satellite images, it is possible to extract flood maps at a high spatial resolution (10 m for Sentinel-1, for example). 

By contrast with water depth measurements at single points, 
the choice of a performance measure for 2D flood maps is yet unclear. Indeed, in a data assimilation framework the simulated water depths are typically compared with the observation in a $L_{2}$ space and considering Gaussian additive errors \cite{hostache2010assimilation,cao2019modified}. However, reconstructing water depths from satellite flood maps is a highly uncertain procedure \cite{schumann2007high,hostache2009water}. Then, one could prefer to directly compare flood maps without introducing additional errors from the reconstruction step. It is possible to make a hypothesis on the error statistical distribution of the flood maps to find the best metric space where to find optimal simulation. However, finding the error prior distribution is complicated (depending on the topography, vegetation, flood extent identification, etc.), so one may prefer to compare indirectly the properties of various performance measures to noised simulations. Then, for evaluating indirectly a performance measure, the natural idea is to create the following twin experiment: i) Create a reference simulation, ii) Create a sample of noised simulations  reproducing the source of errors (satellite position, correction algorithm, etc.), and iii) Identify the ``distance functions to the reference'' with the best mathematical properties for optimization. Reproducing the identified source of errors is difficult. Furthermore, ranking the performance measures according to their properties is not straightforward. In view of the number of difficult hypotheses, we prefer to use an even more indirect method: the possibility of the ``expert'' to identify the optimal parameterization among an ensemble of simulations by testing different criteria.


Weather Forecasting has pioneered the comparison of spatial fields \cite{jolliffe2012forecast}, developing performance measures mainly for rainfall and temperature fields, but could potentially extend to hydraulics as well \cite{casati2008forecast,gilleland2009intercomparison}. 
Until now, most flood events studies rely on the comparison of simulations and observations projected on the same grid with a grid-to-grid (or pixel-to-pixel) approach \cite{aronica2002assessing,grimaldi2016remote}. These performance measures provide incomplete information about the quality of a simulation, as it compares on a pixel-to-pixel basis without considering location mismatch, spatial structure and fields features \cite{jolliffe2012forecast,casati2008forecast,wilks2011statistical}. For example, a simulation with a similar size of flood extent as the observation might yield a poor performance measure value if it is translated by few pixels \cite{gilleland2009intercomparison}.  Pappenberger et al. \cite{pappenberger2007fuzzy} reviewed performance measures based on pixel-to-pixel comparisons. They highlighted that different performance measures should be used together to evaluate different characteristics of the flood. Stephens et al. \cite{stephens2014problems} also assessed the use of pixel-to-pixel performance measures. They show the inconsistency of the performance measures depending on flood event magnitude or catchment characteristics. Further evaluations of performance measures have been performed in hydrology. Cloket et al. \cite{cloke2008evaluating} proposed a six-criteria approach to select new performance measures. Crochemore et al. \cite{crochemore2015comparing} compared expert judgement and performance measures, underlining that they may not rank the best matching hydrograph between simulations and observations in the same order.  

In the hydraulic community, few works explore the use of novel performance measures or the comparison of performance measures for flood maps on real flood events. Thus, the purpose of this paper is to define a specific methodology adapted to flood maps. Criteria are proposed to compare and rank 
measures based on satellite characteristics, flood processes, and computation time. As explained above, due to the difficulty of modelling and simulating the various sources of spatially distributed observation errors, we resort to such a ``metaverification'' approach \cite{murphy1996finley}. Traditional performance measures based on pixel-to-pixel comparisons are assessed, as well as more complex performance measures based on spatial patterns and structures identification. Moreover, although the methodology of this work is applied to flood maps, it can be used with other distributed spatial fields such as water depths (with adapted performance measures). For numerical illustration, a flood event on the Garonne River in France has been chosen, which has been studied several times for flood applications in the context of parameter identification \cite{besnard2011comparaison,el2019uncertainty,nguyen2022dual}. 

The article is structured as follows. Section 2 motivates the problem and introduces how to illustrate its treatment in the case of a real flood event: the flood 
model to be calibrated, the satellite image available on the study area, and the performance measures are described. Section 3 explores various proposed criteria to select the most informative performance measures. These criteria are tested on the study area and the results are analyzed. These results are discussed in Section 4 and a ranking methodology between the performance measures is proposed before concluding with an example of calibration and detailing the limitations of the methodology. Finally, Section 5 presents a conclusion and perspectives of this work.
\section{Materials}\label{sec:material}

The goal is to evaluate and select performance measures for the comparison of a numerical flood model with satellite images of a real flood event. In the present section, the materials needed for the discussion in the context of floods are described. Indeed, the discussion will be illustrated by the numerical application of our method in the specific case of a flood that occurred in France, for which satellite images are freely available and numerical simulations can be performed by a standard flood model. 

The area of the study case is the Garonne River between Tonneins and La Réole located in the South-West of France (see Figure \ref{fig:studyarea}). It is about 50 km long with a mean width of 250 m under normal flow conditions. The floodplain is 1 to 4 km wide along the channel. 
The floodplain is mainly rural areas used for agricultural purposes and the rest are small patches of vegetated and urban areas. 
Since the nineteenth century (after a major flood in 1875), floodplains have been largely equipped with dikes and levees to protect urban areas and infrastructures from flooding. That area has already been used many times for the study of numerical models of floods: we will build on previous efforts as regards the construction of a numerical model (see below). 

\begin{figure}
    \centering
    \includegraphics[width=0.85\linewidth]{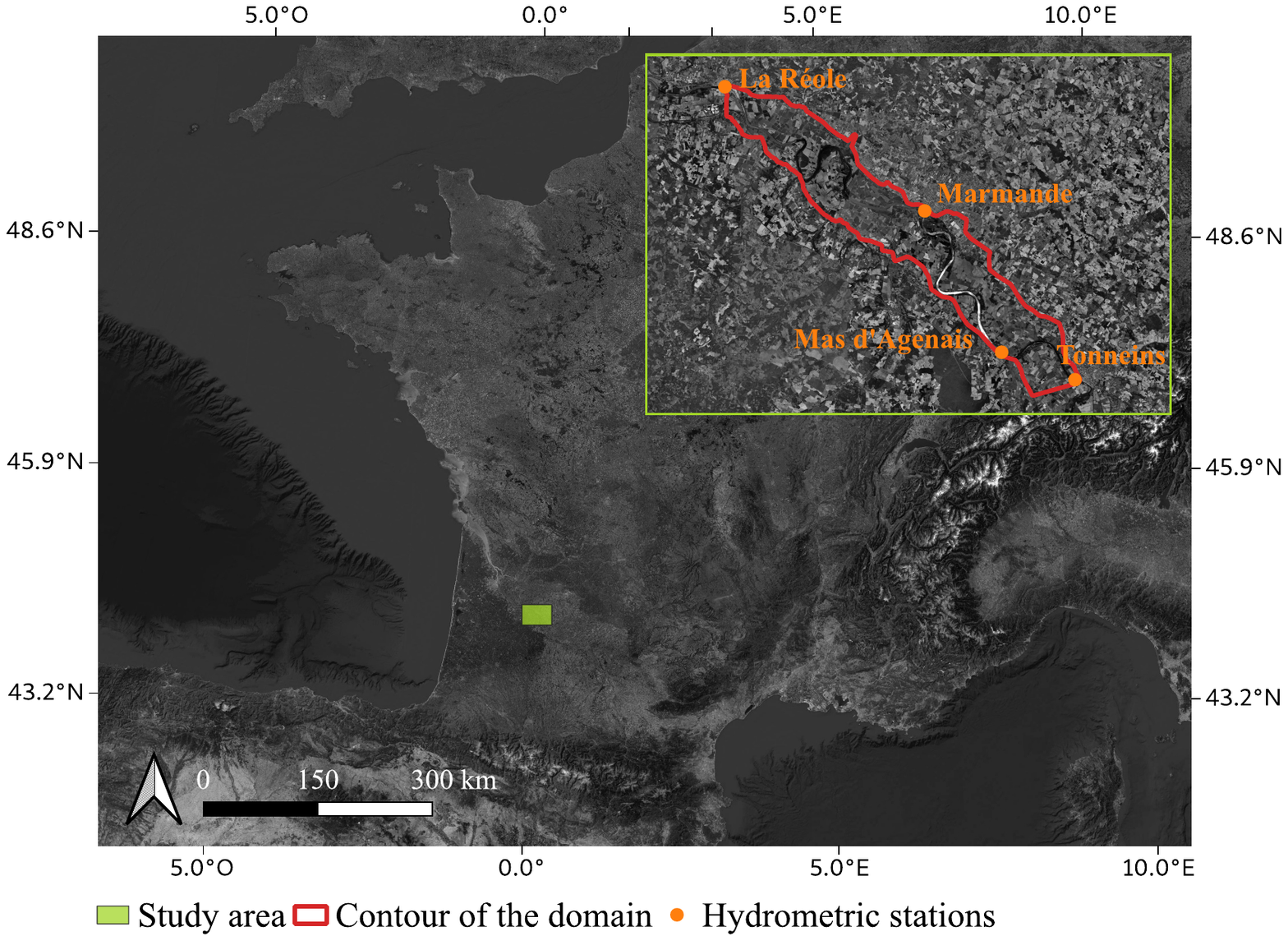}
    \caption{Visualization of the study area on the Garonne River in France.}
    \label{fig:studyarea}
\end{figure}
\subsection{Satellite observation}\label{sec:satelliteobservation}

In February 2021, a major flood occurred over the study area. The flood event was observed twice by the Synthetic Aperture Radar (SAR) Sentinel-1 mission. The time of acquisition of the satellite observations are reported in Figure \ref{fig:discharge}, along with the discharge measured every 15 minutes at three in-situ stations.  All the times indicated in the following are given in Coordinated Universal Time (UTC). 

\begin{figure}

    \centering
    
    \includegraphics[width=0.68\linewidth]{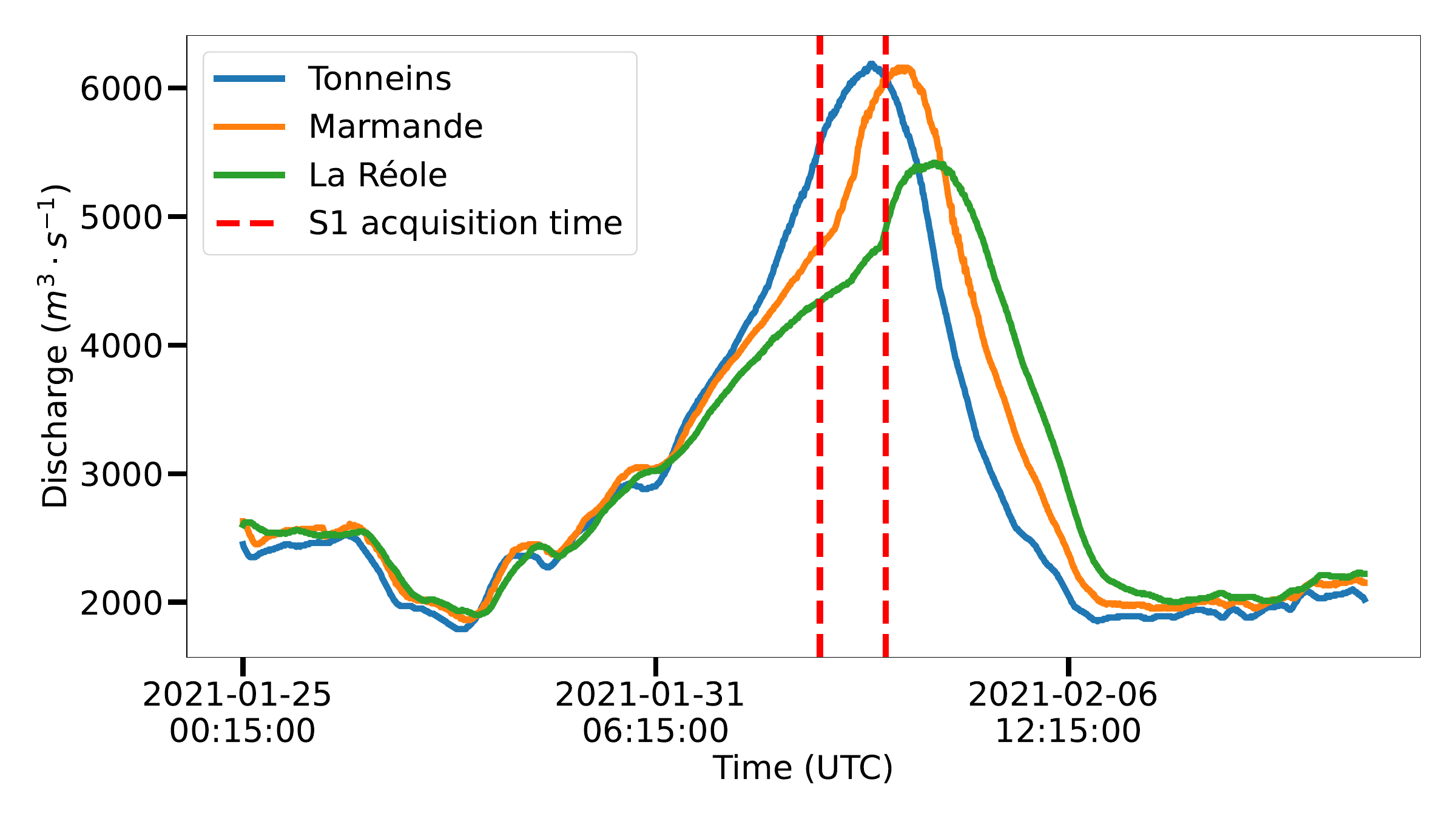}
    
    \label{fig:figb}

  \caption{Measured discharge at the gauging stations (Tonneins, Marmande, and La Réole) and Sentinel-1 acquisition times during February 2021 flood event.}
  \label{fig:discharge}
\end{figure}

For this work, the first satellite observation on February 2, at 5:55 pm is used. 
The image was downloaded from ASF Data Search Vertex (\url{https://search.asf.alaska.edu/}). The observation extracted at ``LEVEL-1 GRD'' is a pre-processed  $N_{x} \times N_{y}$ grid of pixels (involving orbit correction, radiometric calibration, noise reduction, thermal noise removal, terrain correction, and georeferencing) at a resolution of 10 m by 10 m. Here, the dimensions of the image are $N_{x}=2644$ and $N_{y}=2312$.
Then, each pixel is classified as flooded or dry to create a \emph{flood map}. This task can be carried out with multiple extraction procedures \cite{grimaldi2016remote,landuyt2018flood,schumann2023flood}, but this work relies on one method, as it is not the focus of this study. The flood maps are extracted using a local thresholding procedure that distinguishes flooded and dry pixels based on the values of their backscattering (proportional to the energy received over emitted by the satellite on the pixel) denoted as $\sigma^{0}$ . From the histogram of a satellite observation, a threshold $\sigma^{0}_{ref}$ is determined to separate flooded pixels from dry pixels. A more complete description of the principle and of the algorithm can be found in \cite{chini2017hierarchical} and in Appendix A. This method is widely used in flood mapping studies and is considered reliable for the extraction of flood extent \cite{landuyt2018flood,hostache2018near}.

\subsection{Numerical simulation by a hydrodynamic model}

For the simulation of flood events, a numerical solution of the 2D 
Shallow-Water Equations (SWEs) (\ref{eq:swe}-\ref{eq:thirdSWE}) is considered. The SWEs written in a Cartesian coordinate system where gravity acts uniform as $-g\overrightarrow{e_z}$: 
\begin{align}\label{eq:swe}
  \frac{\partial h}{\partial t}+\frac{\partial (hu)}{\partial x}+\frac{\partial (hv)}{\partial y}&=0\\
     \frac{\partial (hu)}{\partial t}+\frac{\partial(hu^{2})}{\partial x}+\frac{\partial (huv)}{\partial y}&=-gh\frac{\partial \eta}{\partial x}+\nabla \cdot (h \, \nu_{e}\nabla u)-\frac{1}{\rho}\tau_{bx},\\ 
          \frac{\partial (hv)}{\partial t}+\frac{\partial(hv^{2})}{\partial y}+\frac{\partial (huv)}{\partial x}&=-gh\frac{\partial \eta}{\partial y}+\nabla \cdot (h \, \nu_{e}\nabla v)-\frac{1}{\rho}\tau_{by}  \label{eq:thirdSWE}
\end{align}
are standardly used to model free-surface flows using the water depth $h\ge 0$ and a depth-averaged velocity $u\overrightarrow{e_x}+v\overrightarrow{e_y}$ as unknown variables functions of $x$, $y$ and time $t\in [0,T)$.
We denote $\eta=h+z_b$ the water surface where $z_b(x,y)$ is the given bottom elevation, $\nu_{e}>0$ a constant viscosity, and
$\tau_{bx}\overrightarrow{e_x}+\tau_{by}\overrightarrow{e_y}$ the bed shear stress function of $h$, $u$, and $v$.

The Manning-Strickler formula \cite{manning1890flow} is used 
for the bed shear stress:
\begin{equation}
   \left\{
    \begin{array}{ll}
\tau_{bx}=\frac{-g\cdot u}{h^{4/3}\cdot K_{s}^{2}}\sqrt{u^{2}+v^{2}} \\[2ex]
\tau_{by}=\frac{-g\cdot v}{h^{4/3}\cdot K_{s}^{2}}\sqrt{u^{2}+v^{2}}
    \end{array}
\right.
,\label{eq:fricT2D}
\end{equation}
where, for application to real floods, 
the spatially-distributed Strickler coefficient $K_{s}$ needs calibrating, as a friction parameter function of $x$, and $y$ \cite{morvan2008concept}. 

TELEMAC-2D (T2D), a module of the open-source hydro-informatics system openTELEMAC (\url{www.opentelemac.org}) \cite{hervouet2007hydrodynamics}, is used in this work to solve (\ref{eq:swe}-\ref{eq:thirdSWE}) complemented by boundary conditions. 
It relies on a (continuous, piecewise-linear) finite element approach, based on an unstructured triangular mesh. The N edge-by-edge scheme and the Positive Streamwise Invariant distributive scheme are used to compute the advection of velocity and water depth, respectively. An important feature for floodplain modelling is the wetting and drying of grid elements, which is considered here through a correction of the free surface gradient \cite{hervouet2007hydrodynamics}.

For the present study, the T2D model uses a 
simplicial mesh made up of $\mathcal{N}=41,416$ nodes and the size of the edges of the grid varies from $40 \, m$ in the channel, to $80 \, m$ around the dikes, and $150 \,m$ in the floodplains (Figure \ref{fig:domain}). To accurately represent singularities, such as dikes, the mesh is constrained along different lines. The bathymetry of the channel was reconstructed with measured bottom elevations of 70 cross sections. The topography in the floodplains is set with IGN (The National Institute of Geographic and Forest Information) topographic maps and aerial photographs. A more complete description of the construction of the model can be found in \cite{besnard2011comparaison}. The model is initialized at $t=0$ with a previous-steady state simulation with an upstream discharge of $2300 \, m^{3}\cdot s^{-1}$. For the boundary conditions, a time-varying hydrograph is imposed upstream at Tonneins (see Figure \ref{fig:discharge} for the flood event under study) and a rating curve linking water depth to discharge is imposed downstream at La Réole. 

\begin{figure}
\includegraphics[width=0.85\textwidth]{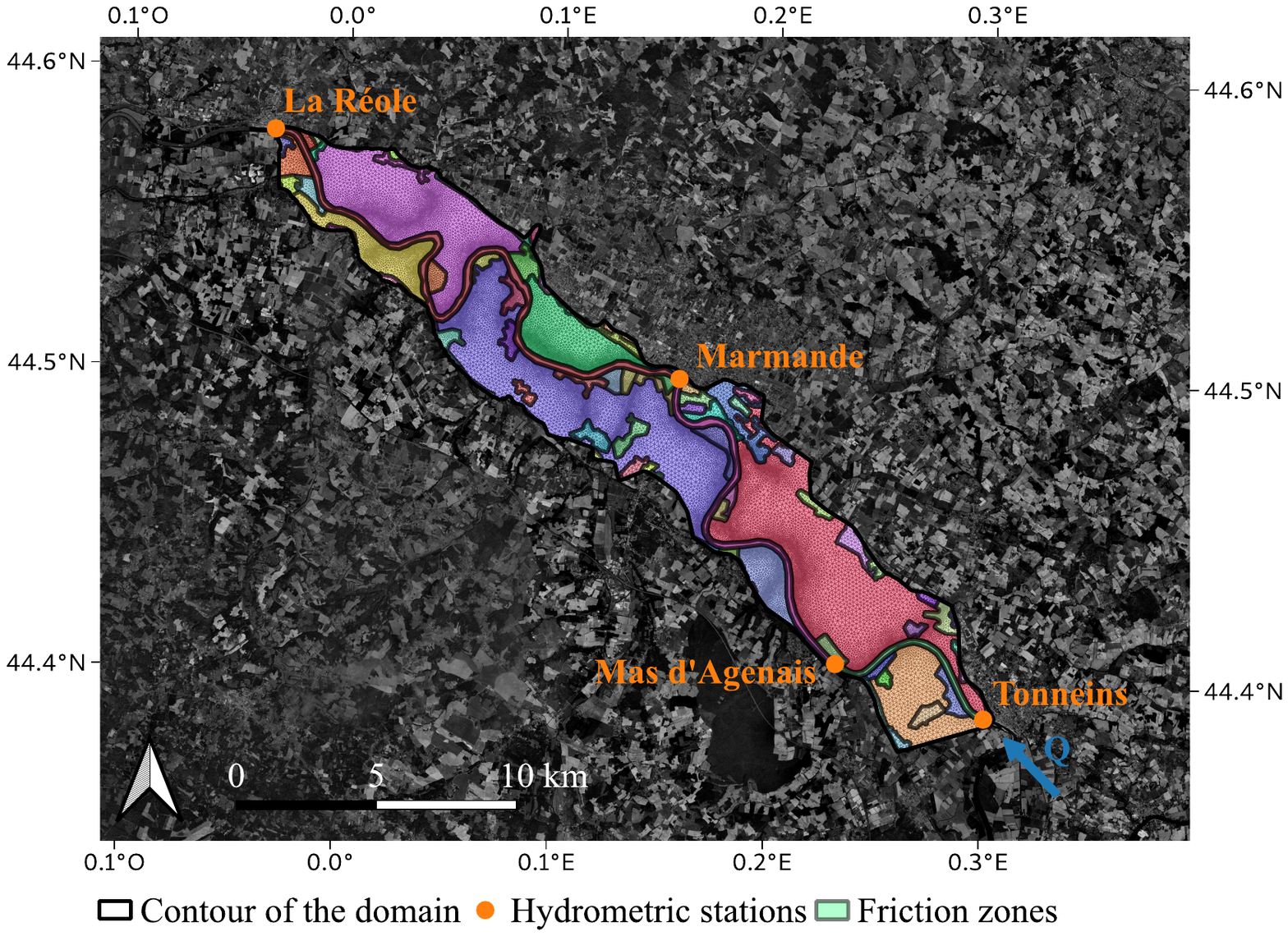}
\caption{Garonne study area and unstructured triangular mesh. The red dots are in-situ gauging stations and the colored subdomains represent the floodplain friction zones.}
\label{fig:domain}
\end{figure}

The friction parameter $K_s$ is piecewise constant over the simulation domain, i.e. with a uniform value 
over 
subdomains. The riverbed is divided into three subdomains based on prior calibration of the model on historical floods: $K_{s}=45 \, m^{1/3}\cdot s^{-1}$ between Tonneins and Mas d’Agenais, $K_{s}=38 \,m^{1/3}\cdot s^{-1}$ between Mas d'Agenais and Marmande, and $K_{s}=40\,m^{1/3}\cdot s^{-1}$ between Marmande and La Réole.
The floodplain is decomposed into 92 subdomains which correspond to areas of
homogeneous land use based on the 2018 Corine Land Cover (CLC) dataset (\url{https://land.copernicus.eu/en/products/corine-land-cover/clc2018}) visualized in Figure \ref{fig:domain}. For each land use class, the a priori range of Strickler values is defined in Table \ref{table:StricklerMinMax}. The range limits are fixed according to expert opinions based mainly on 1D considerations and laboratory experiments \cite{barnes1967roughness,chow1988applied}. A uniform distribution is considered since the shape of the Strickler distribution for each land use is not described in the literature.

\begin{table}
\caption{Range of the friction values for each floodplain land use.}
\label{table:StricklerMinMax}
\centering
 \begin{tabular}{l c}
 \hline
      Land use & Range ($m^{1/3}\cdot s ^{-1}$)\\
      \hline
      Waterbodies & $\mathcal{U}(32.5,37.5)$ \\
      
      Fields and meadows without crops & $\mathcal{U}(17.5, 22.5)$ \\
      Cultivated fields with low vegetation &$\mathcal{U}(15, 20)$ \\
      Cultivated fields with high vegetation & $\mathcal{U}(10, 15)$ \\
      Shrublands and undergrowth areas& $\mathcal{U}(8, 12)$ \\
      Areas of low urbanisation & $\mathcal{U}(8, 10)$ \\
      Highly urbanized areas & $\mathcal{U}(5, 8)$ \\
      \hline
    \end{tabular}
\end{table}

\subsubsection*{Extraction of flood maps from simulations}\label{sec:extract}

The water depth 
is the quantity of interest, output of the simulation. But to be compared with (flood map) observations, it needs processing.
For each simulation, we define a matrix $H\in \mathbb{R}^{N_{x}\times N_{y}}$ such as $H_{i,j}$ equals the value of the finite element water depth at the center of the pixel $(i,j)$.
The latter 
can then be converted into a flood map matrix using a threshold $\tau$ on the water depths. We denote $S^\tau$ a map such that $S^{\tau}_{j,k}=\mathds{1}_{H_{i,j}>\tau}$ for each pixel $(i,j)$ of the observation.
Here, we follow Aronica et al. \cite{aronica2002assessing} and we fix $\tau$ to 10 cm most of the time: 
then we simply rewrite $S^{\tau}$ as $S$.

\subsection{Performance measures}\label{sec:perfmeasure}
The following subsections detail the performance measures used in this work. A number of performance measures are already known to the hydraulic community: mostly those based on pixel-to-pixel comparisons of 2D flood maps \cite{aronica2002assessing,grimaldi2016remote,pappenberger2007fuzzy}. More ``complex'' performance measures based on shape analysis or texture identification are also presented. 28 are evaluated here, which were chosen from the literature based on their 
ease of implementation in the study case context \cite{wilks2011statistical,choi2010survey}. 

In the following, $O,S \in \{0,1\}^{N_x \times N_y}$ refer to flood map matrices from  observation and simulation respectively. We recall the state of each pixel is either dry (0) or flooded (1). The coordinates of the center of the pixels are stored in two matrices $X,Y\in \mathbb{R}^{N_{x}\times N_{y}}$. The observation $O$ was described in Section \ref{sec:satelliteobservation} and the simulated flood maps were described in Section \ref{sec:extract}.  The total number of pixels $N_{x}\times N_{y}$ is denoted by $N$. The term $\mathcal{L}(.,.)$ is used to denote any performance measure.

\subsubsection{Pixel-to-pixel comparison of flood maps}\label{sec:pixel}
Pixel-to-pixel performance measures are ``global'' and compare the number of pixels that match or do not match between a simulation and an observation. There are four possible outcomes that can be represented in a $2\times 2$ confusion matrix (or contingency table), as shown in Table \ref{table:contingency}.  

\begin{table}
\caption{Confusion matrix.}
\label{table:contingency}
\centering
 \begin{tabular}{c c c}
 \hline

 &  Observed flooded & Observed dry  \\

Simulated flooded & TP& FP\\

Simulated dry & FN&TN\\
\hline
\end{tabular}
\end{table}

On the grid points, a hit occurs when a pixel is correctly predicted to be flooded (True Positive ($TP$)) and a correct rejection (True Negative ($TP$)) occurs when it is correctly predicted to be dry. A false alarm (False Positive ($FP$)) and a miss (False Negative ($FN$)) occur when a pixel is simulated as flooded or dry, respectively, while observed as the opposite.

Numerous performance measures combining $TP$, $TN$, $FP$, and $FN$ exist in the literature. The most widely used performance measures are listed in Table \ref{table:performance}. The table reports the range of possible values, the perfect value (for the same flood maps), and a description of each performance measure.

In this work, 18 performance measures listed in Table \ref{table:performance} are evaluated. $F_{\beta}$ is calculated for three values with $\beta \in \{1; 1.5; 2\}$.  Many variants were formulated in the literature to reflect the need to assess classification from different perspectives and fit the specific requirements of different domains. For example, for floods in sensitive areas, minimizing $FN$ (ensuring high recall) might be more important than ensuring high precision by minimizing $FP$ for the decision makers. Other performance measures, such as Matthews-Correlation-Coefficient ($MCC$) or Cohen's Kappa formula ($\kappa$) aim at balancing the variables of the confusion matrix to avoid bias from unbalanced classes. 

\begin{table}[htbp]
\centering
\caption{Most widely used performance measures for pixel-to-pixel comparison based on the confusion matrix adapted from \cite{grimaldi2016remote}.}\label{table:performance}
\resizebox{1\columnwidth}{!}{\begin{tabular}{|l|l|l|}
\hline
\multicolumn{1}{|c|}{\textbf{Name}} & \multicolumn{1}{|c|}{\textbf{Evaluation}} & \multicolumn{1}{|c|}{\textbf{[min, max, perfect value] Description}} \\ \hline

False positive rate (FPR)/Precision (F) & \rule{0pt}{2.5ex}$FP/(FP+TN)$\rule[-1.5ex]{0pt}{0pt}&[0, 1, 0] Proportion of over-prediction of flooded areas \\
\hline

False negative rate (FNR) & \rule{0pt}{2.5ex}$FN/(FN+TP)$\rule[-1.5ex]{0pt}{0pt}&[0, 1, 0] Proportion of over-prediction of dry areas  \\
\hline
True negative rate (TNR) & \rule{0pt}{2.5ex}$TN/(TN+FP)$\rule[-1.5ex]{0pt}{0pt}&[0, 1, 1] Proportion correct of observed dry areas \\
\hline
True positive rate (TPR)/Recall (H) & \rule{0pt}{2.5ex}$TP/(TP+FN)$\rule[-1.5ex]{0pt}{0pt}&[0, 1, 1] Proportion correct of observed flooded areas  \\
\hline
Positive predicted value (PPV) & \rule{0pt}{2.5ex} $TP/(TP+FP)$\rule[-1.5ex]{0pt}{0pt}&[0, 1, 1] Proportion of positive that are true positive \\
\hline
Negative predicted value (NPV) & \rule{0pt}{2.5ex} $TN/(FN+TN)$\rule[-1.5ex]{0pt}{0pt}& [0, 1, 1] Proportion of negative that are true negative\\
\hline
False discovery rate (FDR) & \rule{0pt}{2.5ex}$FP/(FP+TP)$\rule[-1.5ex]{0pt}{0pt}&[0, 1, 0] Proportion of false positive for predicted positive\\
\hline

Peirce skill score (PSS) & \rule{0pt}{2.5ex} H-F \rule[-1.5ex]{0pt}{0pt}&[-1, 1, 1] Maximizing the difference between H and F \\
\hline 

 Accuracy (ACC) & \rule{0pt}{2.5ex}$(TP+TN)/(TP+FP+TN+FN)$\rule[-1.5ex]{0pt}{0pt}& [0, 1, 1] Proportion correct of total
domain area\\
\hline

F$\beta$-score & \rule{0pt}{2.5ex} $(1+\beta^{2})\cdot(H\cdot F)/(\beta^{2}\cdot F+H)$ \rule[-1.5ex]{0pt}{0pt}&[0, 1, 1] Combine precision and recall\\
\hline

Matthews-Correlation-Coefficient (MCC)& \rule{0pt}{4ex}$\displaystyle \frac{TP\cdot TN-FP\cdot FN}{\sqrt{(TP+FP)\cdot(TP+FN)\cdot(TN+FP)\cdot (TN+FN)}}$ \rule[-3ex]{0pt}{0pt}&[-1, 1, 1] Balance of the four categories\\
\hline


Bias (B) & \rule{0pt}{2.5ex}$(TP+FP)/(TP+FN)$\rule[-1.5ex]{0pt}{0pt}&[0, $\infty$, 1] Ratio 
highlighting over/underprediction \\
\hline

Critical Success Index (CSI)& \rule{0pt}{2.5ex}$TP/(TP+FP+FN)$\rule[-1.5ex]{0pt}{0pt}&[0, 1, 1] Adaptation of ACC to focus on flooded areas\\
\hline

$F^{<3>}$ & \rule{0pt}{2.5ex}$(TP-FN)/(TP+FP+FN)$\rule[-1.5ex]{0pt}{0pt}&[-1, 1, 1] Designed to penalize under-prediction of flooded areas\\
\hline
$F^{<4>}$/Measure of fit & \rule{0pt}{2.5ex}$(TP-FP)/(TP+FP+FN)$ \rule[-1.5ex]{0pt}{0pt}&[-1, 1, 1] Designed to penalize over-prediction of flooded areas \\
\hline
Cohen's Kappa formula ($\kappa$)& \rule{0pt}{4ex} $\displaystyle \frac{2\cdot (TP\cdot TN- FN\cdot FP)}{(TP+FP)\cdot (FP+TN) +(TP+FN)\cdot (FN+TN)}$
\rule[-3ex]{0pt}{0pt} 
&[-1, 1, 1] Balance of the four categories\\
\hline
\end{tabular}}
\end{table}

\subsubsection{Geometric comparison of flood maps}
Another possibility to compare 2D flood maps is to compare their geometric properties. These performance measures may provide more information since a flood has a spatial structure. In this subsection, each performance measure is briefly introduced, and further information can be found in the cited references. 


\paragraph{Distance based measures:}

The performance measures introduced below are based on computing distances 
on set of points (point-to-point or between contours). For all these distances, the performance measure value is zero for a perfect match between $O$ and $S$. \\
\underline{Euclidean distance}:\\
\begin{equation}
    d_e(O, S) = \sqrt{\sum_{i=1}^{N_{x}}\sum_{j=1}^{N_{y}}(O_{i,j} - S_{i,j})^2}
.\end{equation}\\
\underline{Hausdorff distance \cite{hausdorff1927mengenlehre}:}\\
The Hausdorff distance $d_{H}$ measures how far apart are two sets of points (here the flooded pixels) computed as:
\begin{equation}
\begin{aligned}
d_{H}(O, S) = \max \{ & \max_{(i,j),S_{i,j}=1} \{ \min_{(k,l),O_{k,l}=1} { \delta((i,j),(k,l))}\}, \\
&\max_{(i,j),O_{i,j}=1} \{ \min_{(k,l),S_{k,l}=1} { \delta((i,j),(k,l))}\}\}
\end{aligned}
,\end{equation}
where \begin{equation}\label{eq:eucledian}
    \delta((i,j),(k,l))=\sqrt{(X_{i,j}-X_{k,l})^{2}+(Y_{i,j}-Y_{k,l})^{2}}. 
\end{equation}\\ 
\underline{Modified Hausdorff distance \cite{dubuisson1994modified}:}\\
The Modified Hausdorff distance $d_{MH}$  is a variation of the Hausdorff distance to reduce the sensitivity of the performance measure to outliers. It is calculated as: 

\begin{equation}
\begin{aligned}
 d_{MH}(O, S) = \max \{ &\frac{1}{|O|}\sum_{(i,j),O_{i,j}=1}  \min_{(k,l),S_{k,l}=1}   \delta((i,j), (k,l)) , \\ & \frac{1}{|S|}\sum_{(k,l),S_{k,l}=1}  \min_{(i,j),O_{i,j}=1} \delta((i,j), (k,l)) \}     
\end{aligned}
,\end{equation}
where $|O|=\sum_{i,j} |O_{i,j}|$ and $|S|=\sum_{i,j} |S_{i,j}|$.

\paragraph{Other performance measures:}
The next performance measures are grouped together, as they are not directly related to computing the distances between \textit{individual} pixels, but are related to the cost of matching whole sets of flooded pixels.\\
\underline{Procrustes analysis \cite{gower1975generalized}:}\\
The Procrustes analysis aims to find an optimal transformation between $O$ and $S$ (including scaling, translation, and rotation) to minimize the sum of square differences between the transformed set of points. First, the analysis begins by standardizing $O$ and $S$ into two matrices $\mathcal{M}_{1}$ and $\mathcal{M}_{2}$ (centered around their origins and $Tr(\mathcal{M}_{i}\cdot (\mathcal{M}_{i})^{T})=1$ with $i\in \{1;2\}$). Then an optimal transformation represented by an orthogonal matrix $Q$ and a scaling factor $c$ is determined using Singular Value Decomposition. The objective is to minimize the quantity $P^{2}$ calculated as:
 \begin{equation}
     P^{2}=d_{e}(\mathcal{M}_{1}-c\mathcal{M}_{2}Q)
 .\end{equation}
 The value of the performance measure between the two sets of points is then given by the smallest value of $P^{2}$. This performance measure is optimal for a value of zero.\\
\underline{Sliced-Wasserstein distance:}\\
We consider $vec(O)_{norm},vec(S)_{norm}\in [0,1]^{N}$ the vectorisation of $O$ and $S$ normalized such that $\|vec(O)_{norm}\|_1 = 1$ and $\|vec(S)_{norm}\|_1 = 1$. 
We define a matrix of cost $C\in \mathbb{R}^{N\times N}$ defined as $C_{i,j}=\delta(i,j)$ representing the cost to move mass from  pixel $i$ to $j$ in $O$ and $S$ respectively (here, $i$ and $j$ refer to two pixel centers while $\delta$ is the usual Euclidean distance in $\mathbb{R}^{2}$ as in Equation \ref{eq:eucledian}).
Then, a possible performance measure is the 2-Wasserstein distance:
\begin{equation}\label{eq:W2}
    W_{2}(O,S)=(\min_{\gamma \in \Pi}<\gamma,C>)^{1/2}=\left(\min_{\gamma \in \Pi}\sum_{i=1}^{N}\sum_{j=1}^{N}\gamma_{i,j}C_{i,j}\right)^{1/2}.
\end{equation}
with $\Pi=\{ A \in [0,1]^{N\times N}  \mid \sum_{i=1}^{N} \sum_{j=1}^{N}A_{i,j}=1,A \mathbf{1}=vec(O)_{norm},A^{T}\mathbf{1}=vec(S)_{norm} \}$ and $\mathbf{1}\in \mathbb{R}^{N}$ the vector of ones. But numerically computing \eqref{eq:W2} is costly.
Here, we consider the Sliced-Wasserstein distance \cite{bonneel2015sliced} as a practical alternative to \eqref{eq:W2}. It is a Monte Carlo approximation of $W_{2}$ obtained from $K$ 1D projections $(r_{k})_{1 \leq k \leq K}\in \mathbb{R}^{N}$ chosen randomly. Applying \eqref{eq:W2} with $<O,r_{k}>$ and $<S,r_{k}>$ instead of $vec(O)_{norm}$ and $vec(S)_{norm}$, the Sliced-Wasserstein distance corresponds to an empirical 
mean with $K=1,000$ 
in this work; it ranges from 0 for a perfect match to $\|C\|_{\infty}$.\\
\underline{Normalized Mutual Information (NMI) \cite{studholme1999overlap}:}\\
The NMI quantifies the amount of information shared between two sets based on the entropy principle:
\begin{equation}
 NMI(O,S)=\frac{H(O)+H(S)}{H(O,S)}
,\end{equation}
where $H(O)=-\sum_{i=1}^{c}p_{i}log(p_{i})$ represents the Shannon entropy for matrix $O$
 with $c$ the number of classes (2 with flooded and dry pixels) and $p_{i}$ the probability of class $i$. \\ $H(O,S)=-\sum_{i=1}^{c}\sum_{j=1}^{c}p_{i,j}log(p_{i,j})$ with $p_{i,j}$ the joint probability between classes $i$ in $O$ and $j$ in $S$. The NMI ranges from 0 to 1 for the optimal value. \\
\underline{Fraction Skill Score \cite{roberts2008scale}:}\\
The Fraction Skill Score (FSS) is based on a neighbour approach where the neighbourhoods of simulated and observed pixels are compared. For every grid point, the number of flooded pixels within a window of size $n$ around it over the total number of pixels in the window ($n^{2}$) is computed.
 From this process, 2D fraction fields are created for the observed and simulated flood maps called, respectively, $O^{n}$ and $S^{n}$. The Mean Square Error for the observed and simulated fractions of a neighbourhood of length n is given by:
\begin{equation}
    MSE_{n}=\frac{1}{N_{x}\cdot N_{y}}\sum_{i=1}^{N_{x}}\sum_{j=1}^{N_{y}}[O^{n}_{i,j}-S^{n}_{i,j}]^{2}
\end{equation}
The worst possible MSE between a simulation and an observation is calculated as:
\begin{equation}
    MSE_{n(ref)}=\frac{1}{N_{x}\cdot N_{y}}\sum_{i=1}^{N_{x}}\sum_{j=1}^{N_{y}}[(O^{n}_{i,j})^{2}+(S^{n}_{i,j})^{2}]
\end{equation}
Then the FSS for each grid level is calculated as:
\begin{equation}
    FSS_{n}=1-\frac{MSE_{n}}{MSE_{n(ref)}}
\end{equation}
In this work, FSS scores are computed for different levels $n \in \{1; 3; 5; 7\}$. The FSS values ranges from 0 to 1 for a perfect match.

\section{Criteria for performance measures comparison}\label{sec:criteria}
Our goal is to choose the most appropriate performance measure to assess flood event simulations against satellite observations.
In this work, the process of ``metaverification'' introduced by Murphy \cite{murphy1996finley} is used. This approach aims to identify performance measures that verify certain criteria or have some properties (e.g., robustness to translation, noise, etc.). Depending on the nature of the problem (floods, temperature field estimation, rainfall analysis, etc.), the criteria may change \cite{jolliffe2012forecast}, so we propose specific criteria to compare flood maps from simulations and satellite observations. The proposed criteria aim at selecting a performance measure that allows a ``robust'' calibration of the friction parameter $K_s$ within a random sample of simulations,
i.e. a performance measure that allows (through a calibration procedure) to practically identify a physical friction parameter field $K_s$ despite limited observation and computational limitations.

The criteria are not ordered by importance. The first criterion quantifies the sensitivity of the performance measures to errors on the \emph{magnitude} of the flood i.e its spatial extent. Then, two criteria test the robustness of the performance measures to displacement and noise errors. For these criteria, we consider a reference flood map (simulated or observed) and generate variations of this flood map. The sensitivity of the performance measure value to the variations of the flood maps are evaluated numerically, relatively to the fixed reference flood map. Finally, a criterion is based on the computational needs to estimate the performance measures. The first and second criteria were inspired from Cloke et al. \cite{cloke2008evaluating} procedure applied to precipitation fields. In this study, they are adapted to the case of flood maps comparison. Our methodology first analyzes each criterion independently. Performance measures that are not rejected by any of the criteria should then be ranked following the procedure in Section \ref{sec:discussion}.

In the sequel, each criterion is first described. It is then discussed based on a numerical evaluation using the performance measures and the study case described in Section \ref{sec:material} with a random sample of 4,000 simulations
generated from 4,000 values of the 92-dimensional parameter field (the floodplain friction). The random parameter values are generated independently using the distributions provided in Table \ref{table:StricklerMinMax}. 
The number of $M=4,000$ models' evaluations is considered satisfactory to obtain converged statistics for all performance measures (see \ref{appendix:MonteCarlo}). 
Each simulation costs 18 hours to compute $T=13 \text{ days}$ of physical time on a CPU Intel(R) Xeon(R) Platinum 8260 @ 2.40GHz. All simulation results (the water depth and velocity fields every hour of physical time) are stored using 4 To.

In the following, we denote as $H(\omega_{m})$ or $S(\omega_{m})$ the simulated water depths or flood maps,  respectively, of one member of the random sample with $m\in \{1; ...; M\}$.

\subsection{Criterion 1: Sensitivity to magnitude errors}
 A good performance measure should be significantly sensitive to changes in the magnitude of our quantity of interest, that is, the extent of the flood. If there are fewer or more flooded pixels in the domain typically because of the sensitivity of the ``flooded'' status of a pixel (link to the minimum $\tau$ in simulations, and to the post-processing in observations), the performance measure should be affected. 
 
 To quantify that sensitivity to variations (``errors'') in the magnitude of the flood extent, following Cloke et al. \cite{cloke2008evaluating}, we propose to generate quantiles within each simulated or observed flood map after varying the main processing parameter in each case (the threshold $\tau$ or $\sigma^0_{ref}$). 
The relative variations can be analyzed for each performance measure as a function of quantiles, possibly after a preliminary statistical analysis (of the relative variations) in the case of samples (like simulated flood maps resulting from a random distribution of the friction parameter). Let us conduct that analysis on both the simulations and the observations independently. 

\subsubsection{Simulated flood maps}

The volume of water in the simulated domain drives the variations in the extent of the flood. By choosing different water depth thresholds $\tau$, we generate variations of the flood extent to evaluate the sensitivity of the performance measure to magnitude errors. For each simulation output $H(\omega_{m})$, flood maps are created with a varying water depth threshold $\tau=h_{l}$ with $l\in \{10;20; ...; 80;90\}$. The levels $h_l$ are defined as empirical quantiles of the distribution of values $H_{i,j}$ with $i,j\in[1, ..., N_{x}]\times [1, ...., N_{y}]$. 
For every simulation $H_{i,j}(\omega_{m})$ can be ordered by increasing values $((H_{k})_{(i,j)}|k=1, ... ,N_{x}\times N_{y})$, then $h_{l}=H_{k=\frac{l}{100}\cdot N_{x}\cdot N_{y}}$. 

For instance, Figure \ref{fig:quantile1050} reports flood maps generated with two thresholds for one of the simulations.

\begin{figure}[htbp]
    \centering
    \includegraphics[width=0.85\linewidth]{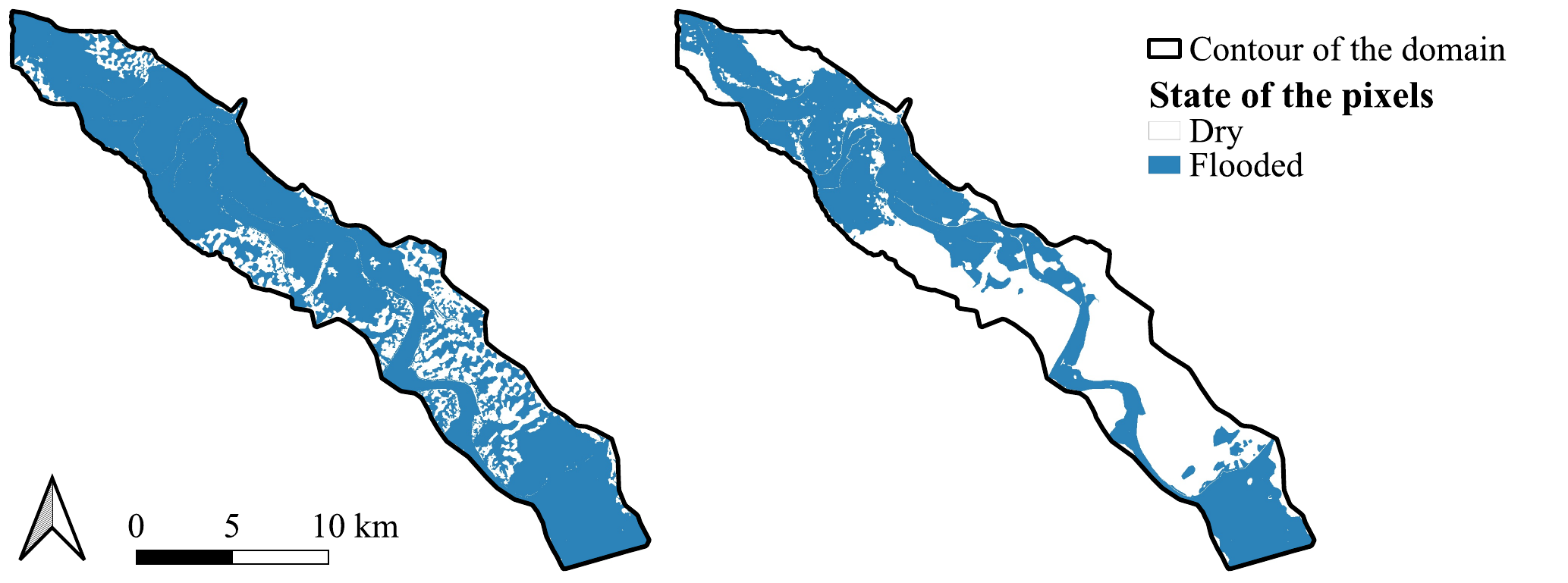}
\caption{Example of flood maps (on the Garonne River) for two thresholds: $h_{20}$ (left), and $h_{50}$ (right).}
\label{fig:quantile1050}
\end{figure}
Then, to check the behaviour of a performance measure $\mathcal{L}$ to changes of flood extent, the map with the quantile $h_{50}$ is chosen as reference and compared with the other quantiles. 
$\mathcal{L}(S^{h_{50}}(\omega_{m}),S^{h_{l}}(\omega_{m}))$ is computed for all simulations $ 1\leq m \leq M$ and all quantiles.
For each quantile, the numerical evaluation of the performance measures can be visualized with box plots representing the range of values for the $M$ simulations. 

The results are visualized in Figure \ref{fig:magnitude}, with red boxes around three specific performance measures to guide the analysis ($FPR$, $ACC$, and $P^{2}$) since they are representative of all the observed behaviours for this criterion.
\begin{figure}[htpb]
    \centering
    \includegraphics[width=0.85\textwidth]{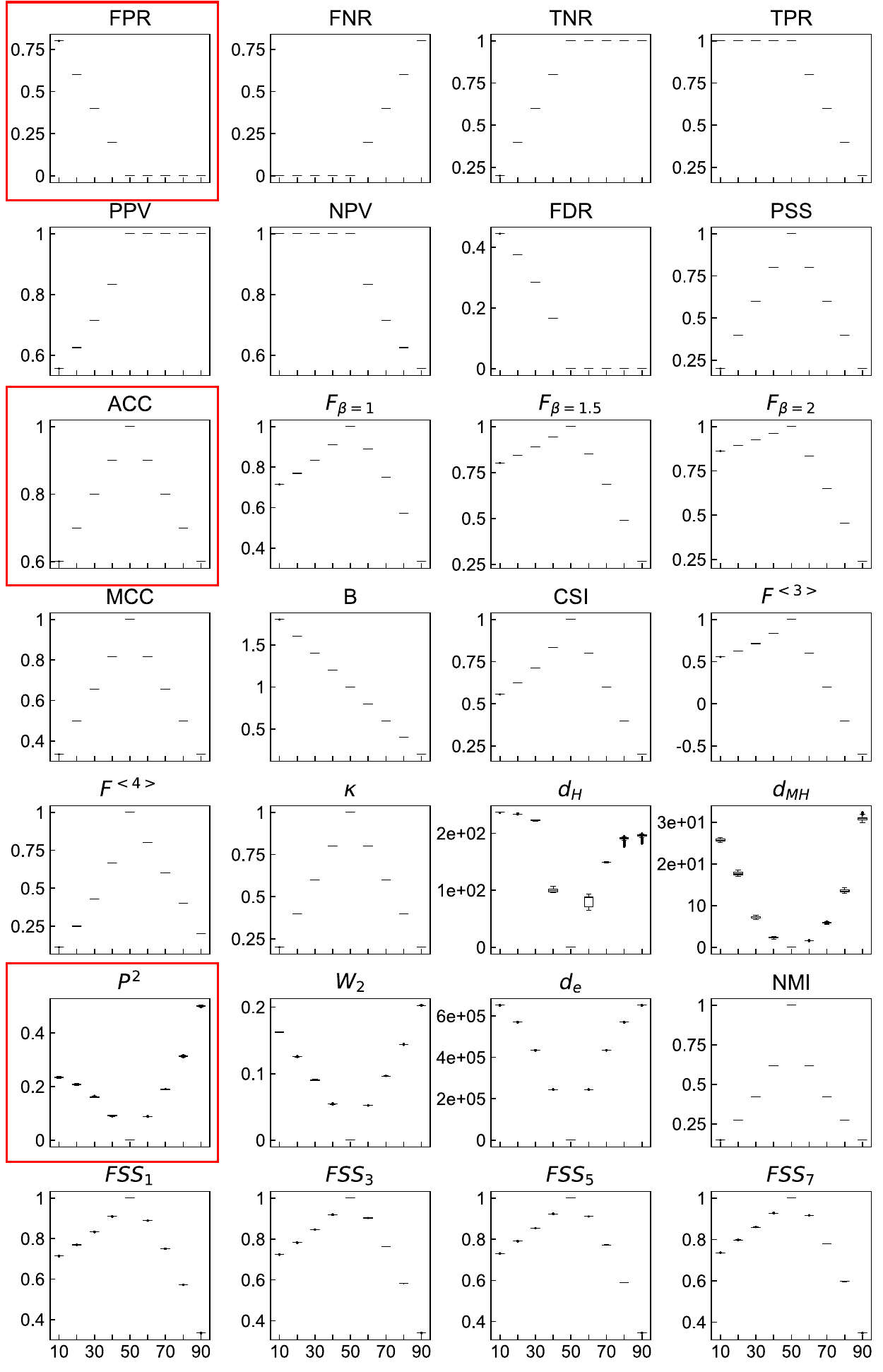}
    \caption{Box plots of $M$ samples $\{\mathcal{L}(S^{h_{50}}(\omega_{m}),S^{h_{l}}(\omega_{m}))\}_{1\leq m \leq M}$ functions of $l\in \{10; ...; 90\}$ for $\mathcal{L}\in \{FPR; ...; FSS_{7}\}$.}
    \label{fig:magnitude}
\end{figure}

To pass the test, the quantity $  |\mathcal{L}(S^{h_{50}}(\omega_{m}),S^{h_{50}}(\omega_{m}))-\mathcal{L}(S^{h_{50}}(\omega_{m}),S^{h_{l}}(\omega_{m}))|  $ is expected to increase when $|l-50|$ increases (the optimal value of each performance measure is considered with $ \mathcal{L}(S^{h_{50}}(\omega_{m}),S^{h_{50}}(\omega_{m}))$ as it is not always equal to zero). For instance, the $ACC$ performance measure verifies this statement since $\mathcal{L}(S^{h_{50}}(\omega_{m}),S^{h_{l}}(\omega_{m}))$ is more distant from its optimal value with increasing errors. $FPR$ does not pass the test as it is insensitive for quantiles $h_{l}>h_{50}$ corresponding to smaller floods compared to the reference map. This result is expected by definition of $FPR$ ($\frac{FP}{FP+TN}$) since the number of false positives is null for flood maps extracted for higher water depth thresholds than the reference maps. Another behaviour is noticed for some performance measures, such as $P^{2}$. The numerical evaluation of $P^{2}$ is not symmetric for negative or positive errors and tends to favour larger floods with higher values.

According to the criterion, some performance measures fail the test. In fact, $FPR$, $FNR$, $TNR$, $TPR$, $PPV$, $NPV$, and $FDR$ are not sensitive to increasing errors on the flood extent for larger or smaller floods compared to the reference optimum. These results align with the performance measure's definitions, which aim to minimize errors for either flooded or dry areas, but not both simultaneously. Their definitions indicate their inability to meet the criterion, yet this criterion aids in analyzing more complex performance measures that cannot be directly interpreted from their definitions. The other performance measures are sensitive to both smaller and larger floods. It can be considered that they meet the criterion. However, as $P^{2}$, other performance measures are unbalanced, being more sensitive to the magnitude test for smaller or larger floods such as $F_{\beta}$, $CSI$, $F^{<3>}$, $P^{2}$ and $FSS$. 
 
\subsubsection{Observed flood maps}
\label{sec:observed}

The observed flood maps may vary depending on the extraction procedure parameters. In this work, the identification of $\sigma^{0}_{ref}$ is subject to uncertainties due to measurement and post-processing errors. To evaluate the behaviours of the performance measures with these uncertainties, flood maps with other thresholds around $\sigma^{0}_{ref}$ are created.
By changing the threshold value, flood maps with more or less flooded pixels are created (similarly to the magnitude errors for the simulated fields). The impact of the threshold error is illustrated in Figure \ref{fig:Gaussian}. Ten flood maps denoted as ($O^{\sigma^{0}_{i}})_{-5 \leq i \leq 5}$ with threshold values $\sigma^{0}_{i} \in \{0.95\sigma^{0}_{ref}, ...,0.99\sigma^{0}_{ref}, \sigma^{0}_{ref}, 1.01\sigma^{0}_{ref}, ..., 1.05\sigma^{0}_{ref} \}$ are created and compared with the reference flood map. Flood maps with the lowest errors to the reference are expected to have performance measure values closer to the reference, similarly to the previous test on the simulated fields. 

\begin{figure}[htbp]
    \centering
    \includegraphics[width=0.85\textwidth]{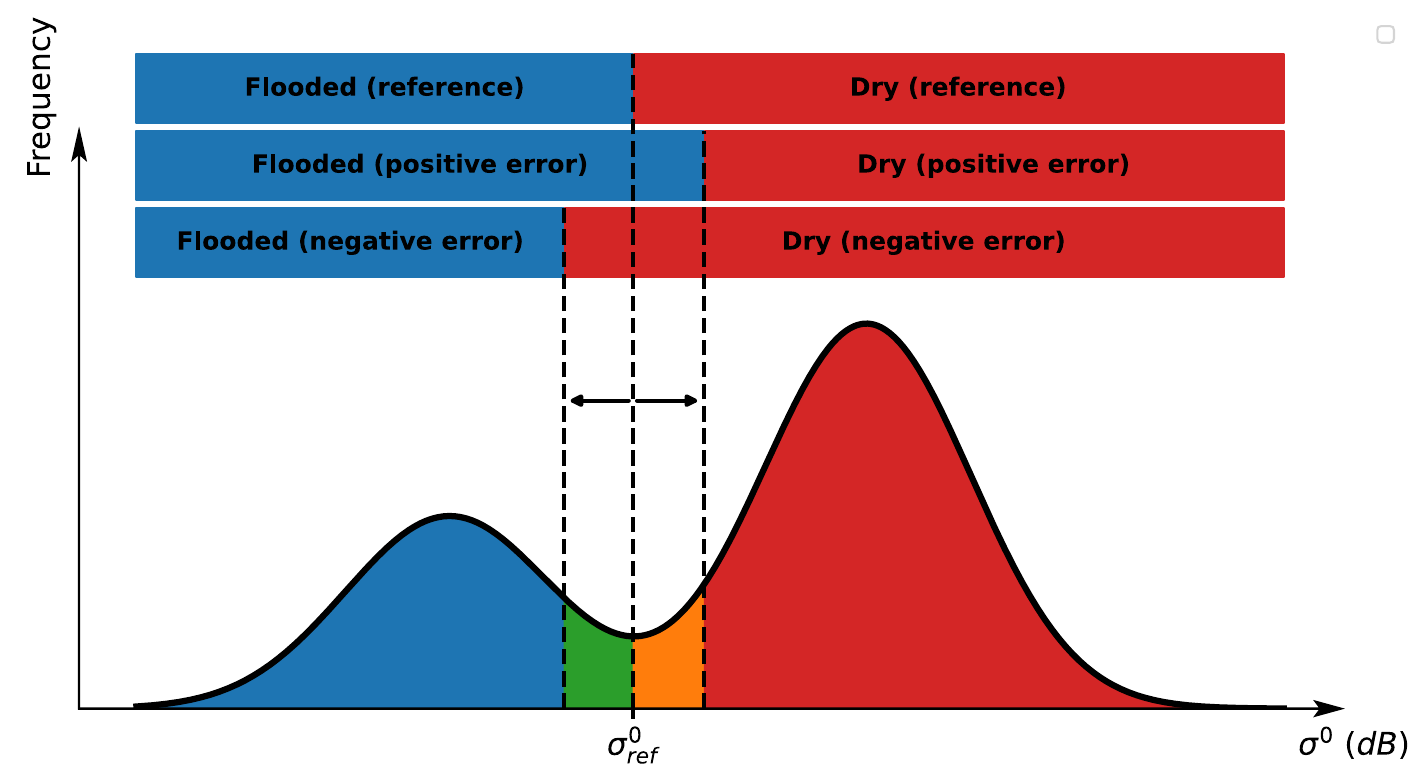}
    \caption{Illustration of the creation of observed flood maps with errors on the threshold identification on bimodal Gaussian distributions of $\sigma^{0}$.}
    \label{fig:Gaussian}
\end{figure}

The analysis of this criterion on the observation described in Section \ref{sec:satelliteobservation} is visualized in Figure \ref{fig:threshold}. The behaviours to threshold errors are consistent with those observed for the criterion on the simulated fields. Similarly to the test on the simulated fields, threshold errors correspond to magnitude errors.
Indeed, the flood extent in the observed flood maps is smaller than the reference map when a negative threshold is applied and larger when a positive threshold is applied. A summary of the magnitude errors criterion is provided in Tables \ref{table:recapmag1} and \ref{table:recapmag2}. 

In the following, the superscript $\tau$ is fixed to 10 cm, so that $S^\tau=S$, and any additional superscript on $S$ is used to denote other transformations of the simulated flood maps.

\begin{figure}[htpb]
    \centering
    \includegraphics[width=0.85\textwidth]{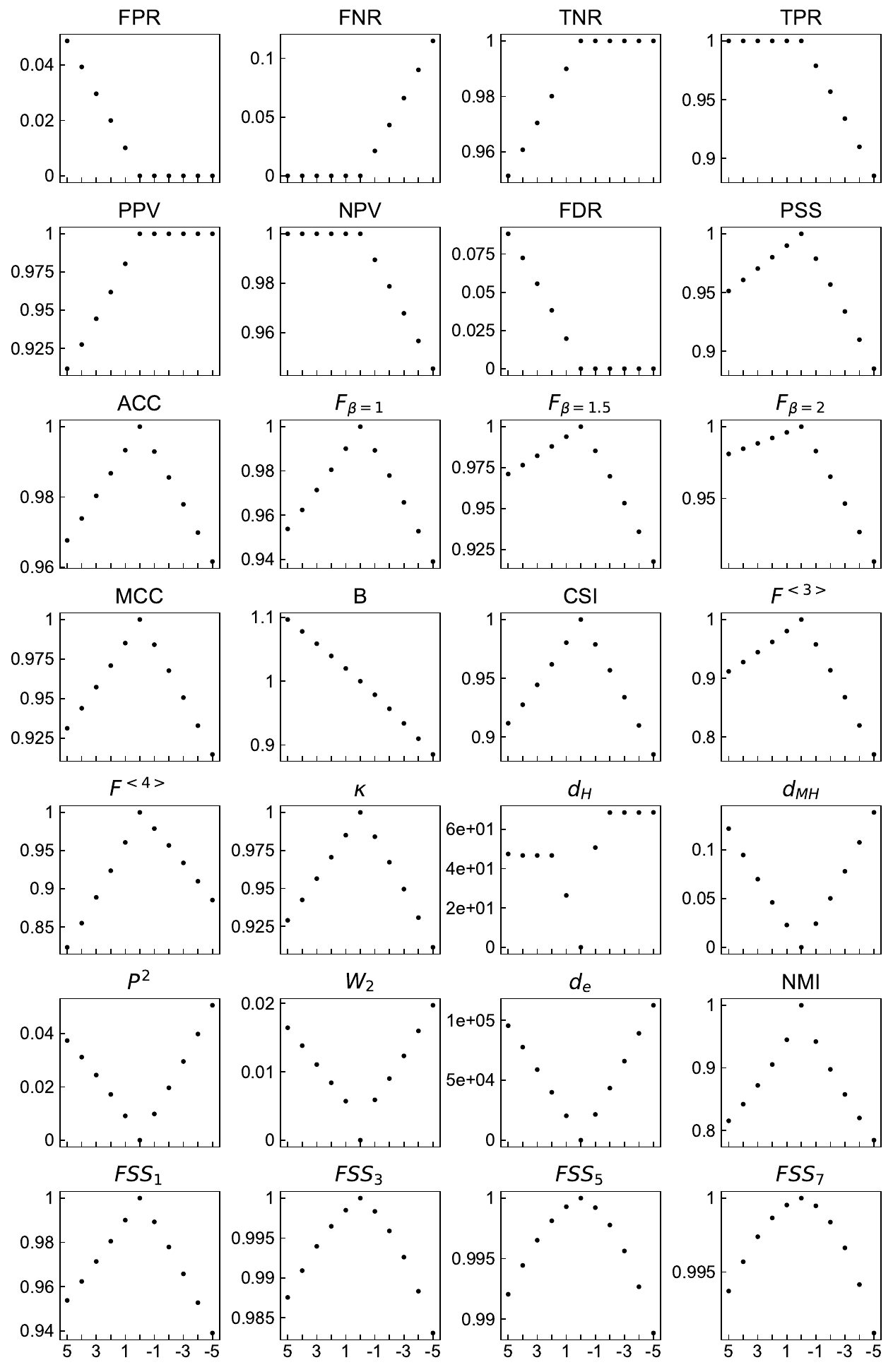}
\caption{Performance measure value $\mathcal{L}(O^{\sigma^{0}_{0}},O^{\sigma^{0}_{i}})$ functions of $i\in \{-5;-4; ...;4;5\}$ for $\mathcal{L}\in \{FPR; ...; FSS_{7}\}$.}
    \label{fig:threshold}
\end{figure}

\subsection{Criterion 2: Sensitivity to displacement errors}\label{sec:displacement}

\subsubsection{Translation errors}\label{sec:translation}
SAR observations can be subject to translation errors in space. A good performance measure should be significantly sensitive to such errors. To quantify the latter sensitivity, we generate translation errors on flood maps in various directions. Each flood map $S(\omega_{m})$ is translated by one to three pixels in three directions (one north (1N), one west (1W), one north-west (1NW), two north (2N), two west (2W), two north-west (2NW), three north (3N), three west (3W), and three north-west (3NW)). Longer distances are not considered, as they do not represent probable translations from geolocation errors or from the extraction of flood maps \cite{schumann2008estimating}. For a grid of pixels with a resolution of 10 m, center-to-center distances between the reference maps and the translated maps range from 10 m to 42 m (three pixels in the north-west direction). The translated fields are denoted as $S^{t}(\omega_{m})$ with $t\in \{0;1N;1W;1NW;2N;2W;2NW;3N;3W;3NW\}$, and $S^{0}(\omega_{m})=S(\omega_{m})$. The quantity $\mathcal{L}(S(\omega_{m}),S^{t}(\omega_{m}))$ is computed for all simulations $ 1\leq m \leq M$ and all translations. The evaluation of the performance measure for a translated field against the reference field should deviate from the optimal value of the performance measure with increasing distance.

The results of this process on the study case are visualized in Figure \ref{fig:shift}.
\begin{figure}[htbp]
    \centering
    \includegraphics[width=0.85\textwidth]{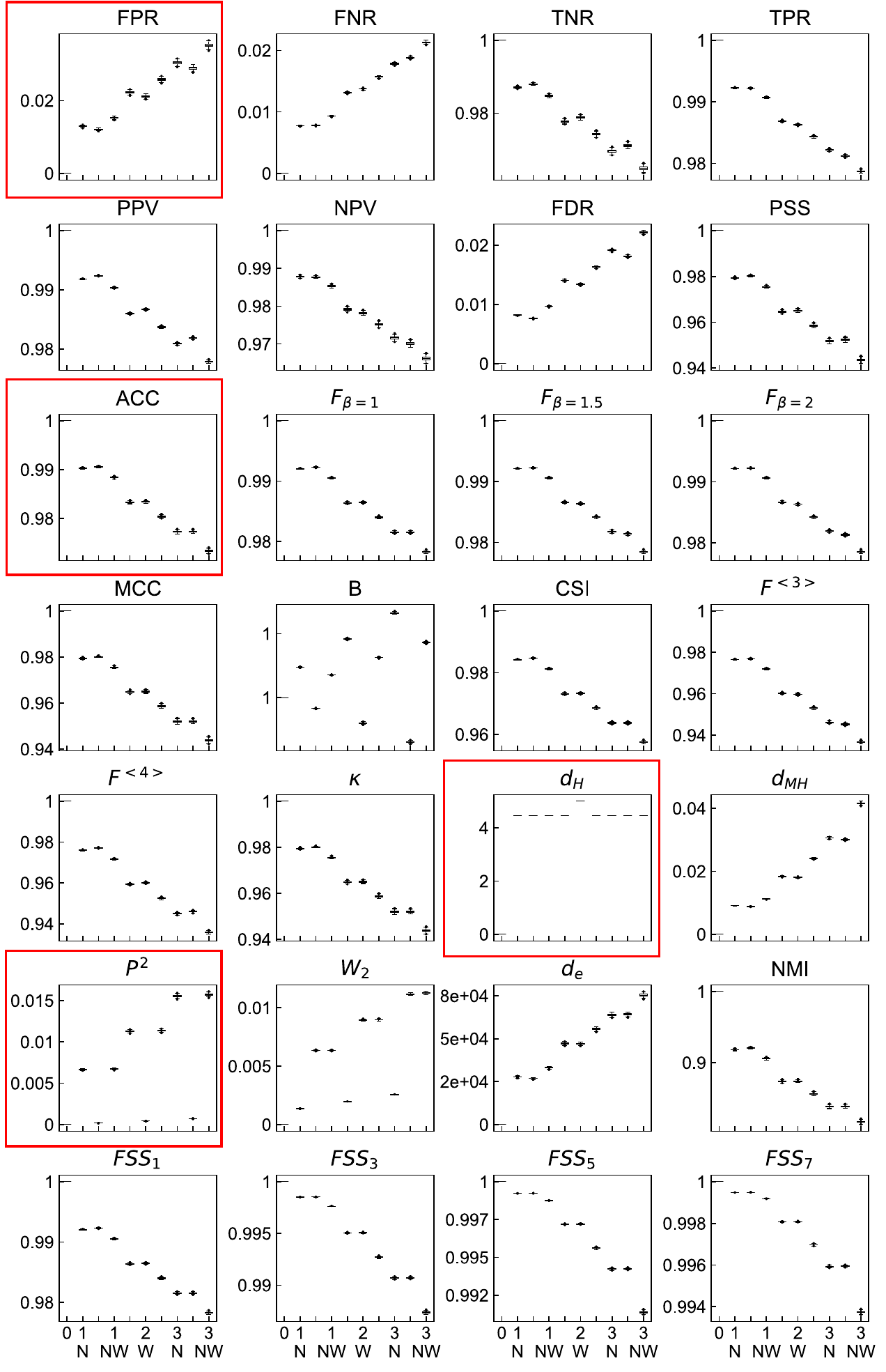}
    \caption{Box plots of $M$ samples $\{\mathcal{L}(S(\omega_{m}),S^{t}(\omega_{m}))\}_{1\leq m \leq M}$ functions of $t\in \{0;1N;1W; ...; 3W;3NW\}$ for $\mathcal{L}\in \{FPR; ...; FSS_{7}\}$.}
    \label{fig:shift}
\end{figure}
Special attention is paid to $FPR$, $ACC$, $P^{2}$ and $d_{H}$ surrounded by red boxes that represent all the possible behaviours of the translation criterion. The box plots are ordered by increasing distance from the reference simulated fields. Then, a performance should pass the test if by visual inspection the trend is monotonous towards decreasing values with respect to the reference. 
While $FPR$ and $ACC$ linearly deviate from optimal values with distance, $d_{H}$ and $P^{2}$ show insensitivity to certain translations. Indeed, $d_{H}$ has a constant value for any translation in the experiment and $P^{2}$ value does not deteriorate for translations in the west direction.  

Most performance measures (such as $FPR$ and $ACC$) respect the criteria with better scores for shorter distances, with three visible steps depending on whether they are translated by one, two, or three pixels. $W_{2}$, $d_{H}$ and $P^{2}$ also fail the criterion to translation errors because of their non-monotonous behaviours with increasing/decreasing distances.

\subsubsection{Rotation errors}\label{sec:rotation}
The observation may also contain rotational inaccuracies due to small changes in the sight angle and distortion errors \cite{sansosti2006geometrical}. The simulated flood maps are rotated with different angles and compared to the reference flood maps. The rotation is performed by rotating counterclockwise around the center. The coordinates of the rotated field are calculated as follows:
\begin{equation}
  \begin{bmatrix}
    (X_{rotated})_{i,j} \\
   (Y_{rotated})_{i,j}
\end{bmatrix}=  \begin{bmatrix}
     cos(\theta)& -sin(\theta)\\
     sin(\theta) & cos(\theta)
\end{bmatrix} \begin{bmatrix}
    X_{i,j} \\
    Y_{i,j}
\end{bmatrix},
\end{equation}
where $\theta$ is the angle of rotation. 
The simulated flood maps are rotated arbitrarily by $\theta \in \{-10^{\circ},-5^{\circ},-2^{\circ},-1^{\circ},0^{\circ},1^{\circ},2^{\circ},5^{\circ},10^{\circ}\}$ and the rotated fields are denoted as $S^{\theta}(\omega_{m})$ with $S^{0^{\circ}}(\omega_{m})=S(\omega_{m})$. Then, similarly to the translation displacement, $\mathcal{L}$ is calculated for each configuration against the reference simulated flood maps. The evaluation of the performance measure for a rotated field should deviate from the optimal value with increasing angles.

The results of these evaluations are visualized in Figure \ref{fig:rotation}, with $B$ and $W_{2}$ surrounded by red boxes representing all the behaviours of the test. The Sliced-Wasserstein distance $W_{2}$ meets the criterion, since the value of $\mathcal{L}(S(\omega_{m}),S^{\theta}(\omega_{m}))$ deviates from its optimal value of zero for increasing rotation angles. All performance measures have similar behaviours to the increase of rotation errors except for the bias $B$. Indeed, $\mathcal{L}(S(\omega_{m}),S^{\theta}(\omega_{m})$) is not monotonous with increasing rotation errors as it oscillates around its optimal value 1 for a rotation angle of $1^{\circ}$ to $5^{\circ}$. 

\begin{figure}
    \centering
    \includegraphics[width=0.85\textwidth]{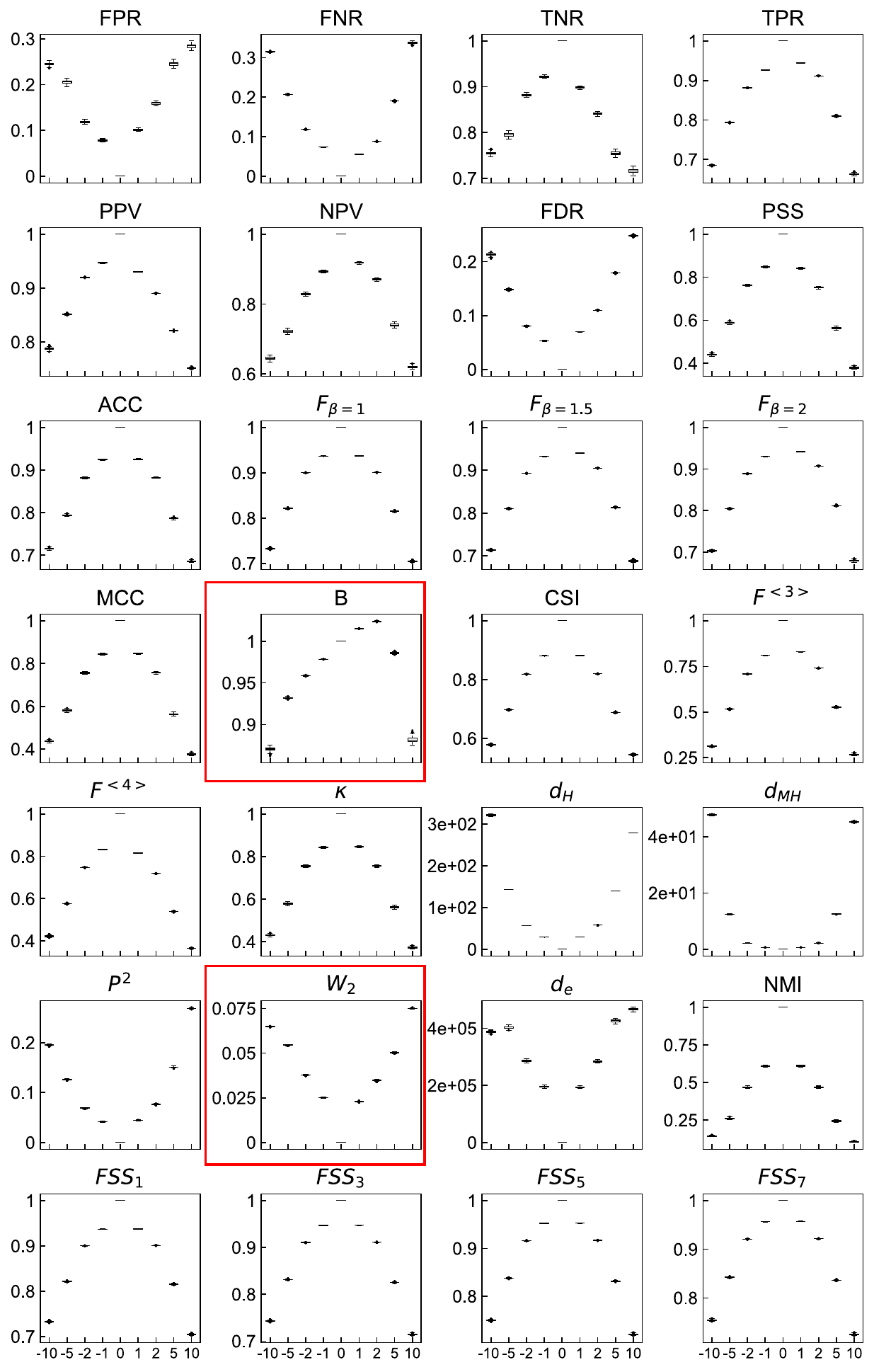}
    \caption{Box plots of $M$ samples $\{\mathcal{L}(S(\omega_{m}),S^{\theta}(\omega_{m}))\}_{1\leq m \leq M}$ functions of $\theta \in \{-10^{\circ},-5^{\circ},...,0^{\circ},...,5^{\circ},10^{\circ}\}$ for $\mathcal{L}\in \{FPR; ...; FSS_{7}\}$.}
    \label{fig:rotation}
\end{figure}

$d_{H}$, $P_{2}$, and $W_{2}$ are rejected by the criterion for translation errors and $B$ for rotation errors. A summary of the translation and rotation criteria is provided in Tables \ref{table:recapmag1} and \ref{table:recapmag2}.

\subsection{Criterion 3: Sensitivity to noise}\label{sec:noise}


A good performance should naturally be sensitive to the noise spatially-distributed within the observation (that is within the satellite images), before or after its processing (as an  extracted flood map here). To evaluate such sensitivity 
Pappenberger et al. \cite{pappenberger2007fuzzy} proposes to generate noised 
using Sequential Gaussian Simulations (SGS) with various spatial correlations based on a variogram computed on the observation. However, SGS is too costly to compute for a large number of pixels (more than 6,113,000 here) because of the variogram modelling and kriging steps. 
Another possibility is to introduce noise with a Gaussian Filter. Gaussian filtering aims to smooth an image by replacing each pixel value with a linear combination of its neighbours. Practically, it involves performing a convolution between an input matrix and a convolution kernel. An illustration is given in Figure \ref{fig:noisekernel}, with a kernel of size 3.

\begin{figure}[htbp]
    \centering
    \includegraphics[width=0.68\textwidth,trim={0cm 0.1cm 0cm 0.1cm},clip]{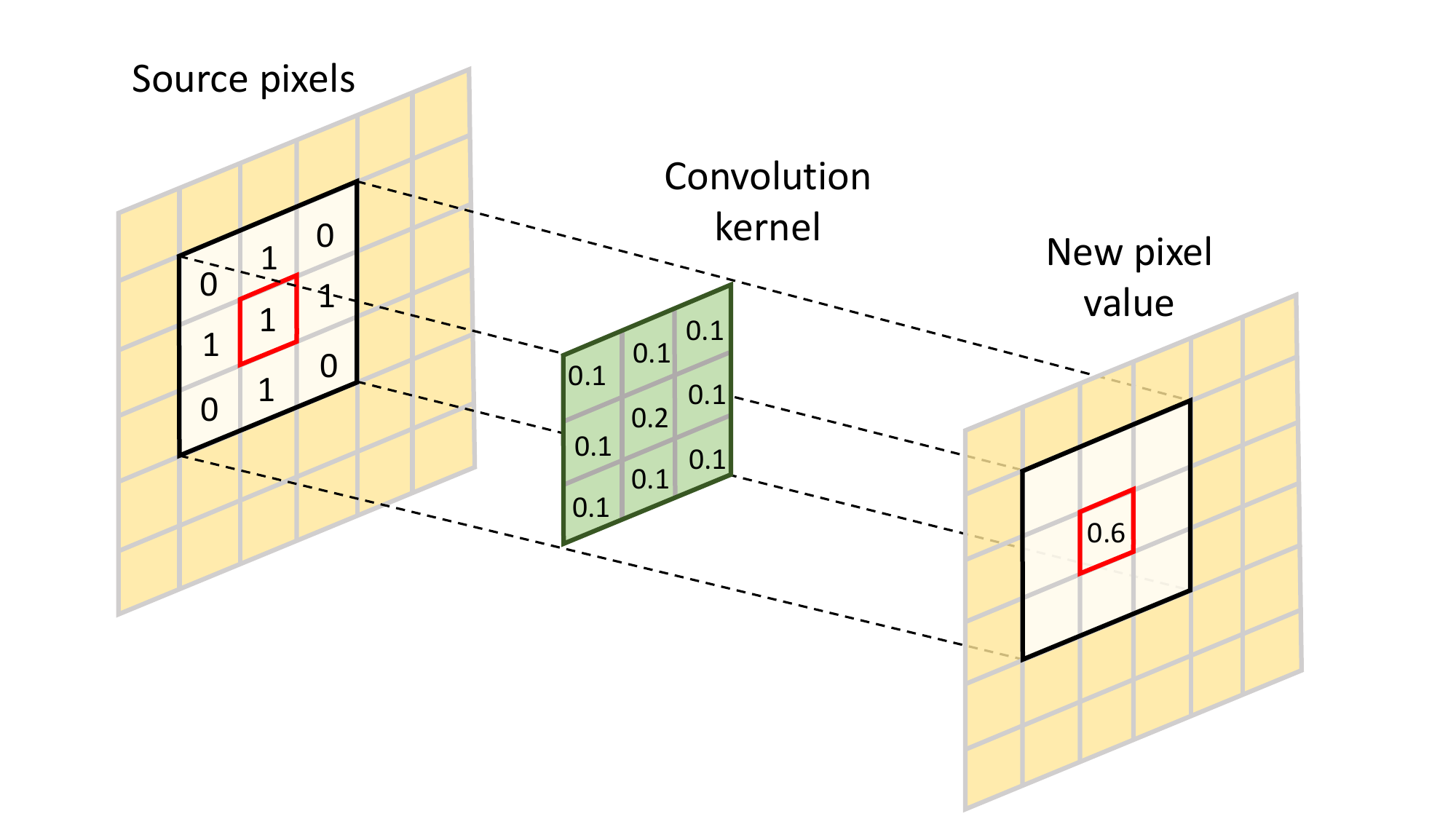}
    \caption{Example of matrix convolution by a convolution kernel.}
    \label{fig:noisekernel}
\end{figure}

For a Gaussian filter, the kernel is computed as follows:

\begin{equation}
   G(X_{i,j}, Y_{i,j}) = \frac{1}{2 \pi \sigma^2} e^{-\frac{X_{i,j}^2 + Y_{i,j}^2}{2 \sigma^2}} 
, \end{equation}

where $\sigma$ is the standard deviation. The kernel matrix is then normalized so that the sum equals one.

The reference observed flood map is noised for different kernel sizes (3, 5, 7, and 9) with $\sigma \in \{0.1, 0.2, \ldots, 2.9, 3.0\}$. After the convolution, pixels with a value above 0.5 are considered to be flooded, and otherwise dry. The noised observations denoted as $O^{\sigma}$ are compared to the reference computing $\mathcal{L}(O,O^{\sigma})$. 

The results are analyzed for a kernel size of 9 with $\sigma \in \{0.1,0.2,...,2.9,3.0\}$ in Figure \ref{fig:noise}. A special attention in the analysis is given to $FPR$, $B$, $d_{H}$ and $P^{2}$ highlighted with red boxes in the Figure as they represent all the observed behaviours. $P^{2}$ shows a strictly monotone trend with respect to the standard deviation. This is the expected behaviour, since higher standard deviations mean that the observations are more noised (more smoothing on the pixels values) with regard to the reference. $FPR$ is insensitive for $\sigma$ lower than 0.5. Then it fails the noise criterion, as it cannot detect the optimal maps (with the lowest noise) due to identifiability issues in these regions. The same issue is raised for $d_{H}$ with jump discontinuities and identifiability problems at the different steps. Another phenomenon is observed for $B$, the performance is not monotone with increasing standard deviation, so maps with different noise levels exhibit the same value, which is not advisable. 

\begin{figure}[htbp]
    \centering
    \includegraphics[width=0.85\textwidth]{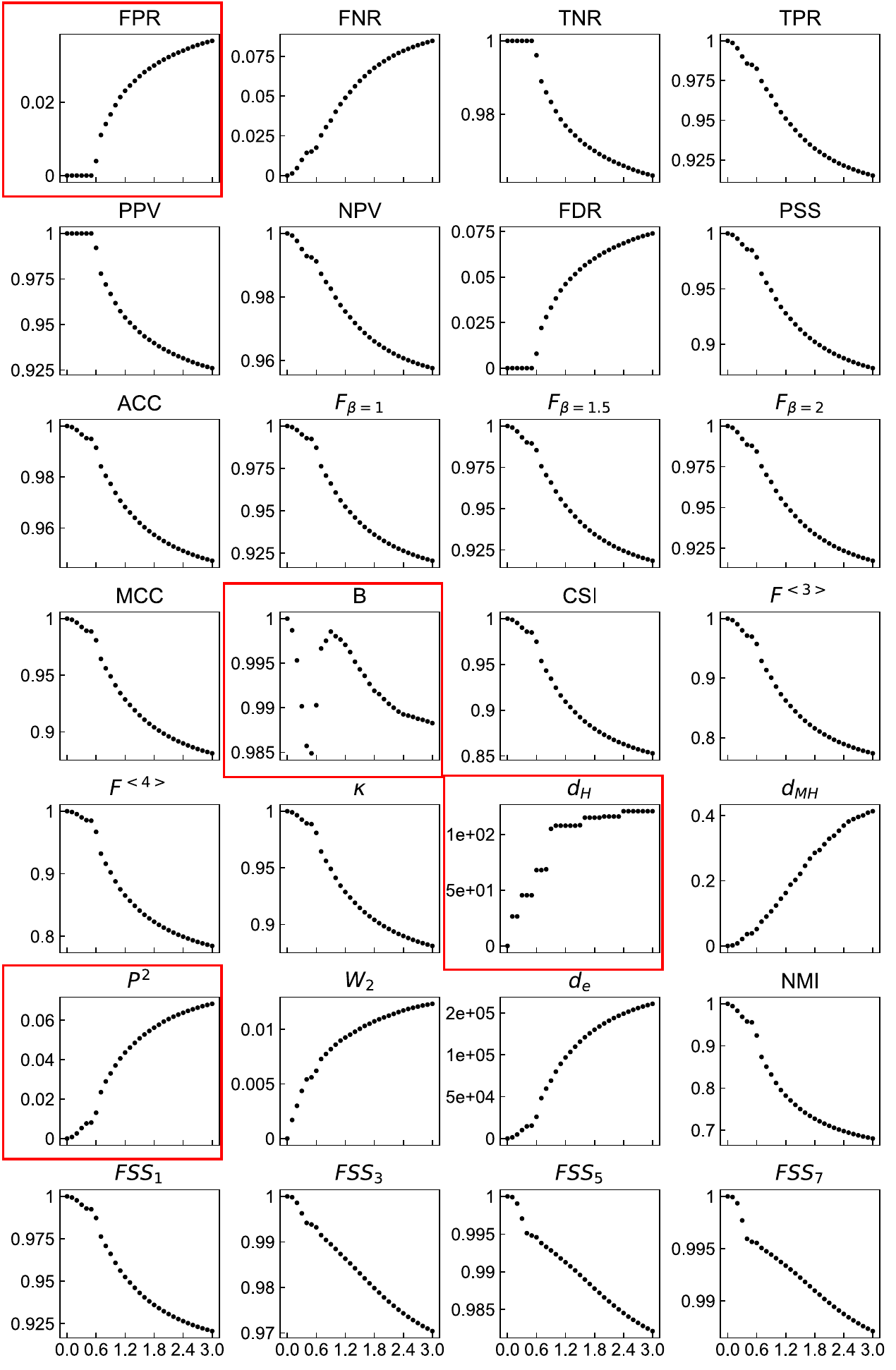}
\caption{Performance measure value $\mathcal{L}(O,O^{\sigma})$ functions of $\sigma\in \{0;0.1; ...;2.9;3\}$ for $\mathcal{L}\in \{FPR; ...; FSS_{7}\}$.}
    \label{fig:noise}
\end{figure}

For this criterion, $FPR$, $TNR$, $PPV$, $FDR$ are rejected because they are not sensitive to small noise levels, $B$ is rejected because it is not monotonous, and $d_{H}$ for the jump discontinuities. 
A summary of the noise criterion is provided in Tables \ref{table:recapmag1} and \ref{table:recapmag2}.

\subsection{Criterion 4: Computation time}

The cost of computing a performance measure could be prohibitive, so, depending on the available computational capabilities, the user may be interested in cheaper implementations.
For each call to $\mathcal{L}$ the computation time is evaluated. The preprocessing of the images, such as projections or flood mapping is not taken into account for the computation time. The selection of a performance measure may depend on the implementation efficiency of the criteria by the developers. 

In Table \ref{table:time}, the computation time of $\mathcal{L}$ between two flood maps $O$ and $S$ is evaluated. The performance measures are calculated on Intel(R) Xeon(R) Platinum 8260 CPU @ 2.40GHz (Cascade Lake) processors. 
\begin{table}[htbp]
\label{table:time}
\caption{Computation time of the performance measure between two flood maps.}
\centering
\begin{tabular}{|c|c|c|c|c|c|c|c|c|c|c|c|c|c|c|c|c|c|c|c|}
    \hline
 & \textbf{Pixel-to-pixel} & \textbf{HD} &\textbf{MH}   & $\mathbf{P^{2}}$ & \textbf{$\mathbf{W_{2}}$} &\textbf{$\mathbf{d_{e}}$} & \textbf{NMI} &\textbf{FSS1-7}  \\
    \hline
Time (s)  &0.09&3.00&2.88&16.32 & 2.38&0.05&0.67&835 \\
    \hline
\end{tabular}
\end{table}

The performance measures based on a confusion matrix and Euclidean distance are fast to compute as they are based on a simple comparison element by element of matrices. For the other performance measures, the cost differs between them. For instance $FSS$ performance measures appear to be costly to calibrate a model in an iterative optimization approach. As for the others, the cost does not seem prohibitive, except perhaps $P^{2}$ for special applications. In this study, $FSS$ are considered to not meet the criterion and partially $P^{2}$, but it could be due to an inefficient optimization of the implementation of the performance measures.   




\section{Discussion}\label{sec:discussion}

This section aims to sum up the performance measures selected from the proposed criteria.
After the selection phase from the metaverification process, correlations between performance measures are evaluated, and the performance measures are ranked according to their robustness to noise. Then, the best performance measures for this test case are used to show an example of possible calibration of the model, and the different limitations of the methodology are discussed. 

\subsection{Summary of the criteria and ranking}
\subsubsection{Summary of the selection phase}

Performance measures were evaluated for each criterion and summarized in Tables \ref{table:recapmag1} and \ref{table:recapmag2}. The number of performance measures have been reduced from 28 initially to 8 (21 have been rejected for at least one of the criterion). The performance measures meeting all criteria are: $PSS$, $ACC$, $MCC$, $F^{<4>}$, $ \kappa$, $d_{MH}$, $d_{e}$ and $NMI$. The number of adapted performance measures is lowered, but for calibrating or validating simulations with observations, the ideal is to have the fewest number of performance measures even in a multi-objective algorithm. Firstly, one approach is to identify correlations between performance measures to remove redundant ones, while a second approach involves ranking them based on a specific methodology. Both methods are described in the following.
\begin{table}
  \centering
  \caption{Summary of the metaverification process for pixel-to-pixel comparisons. $\times$ are rejected by the criterion, $\checkmark$ are accepted by the criterion, and $\sim$ are in between.}
 
    \resizebox{\linewidth}{!}{\begin{tabular}{|c|c|c|c|c|c|c|c|c|c|c|c|c|c|c|c|c|c|c|}
      \hline
     & \textbf{FPR} & \textbf{FNR} & \textbf{TNR} & \textbf{TPR} & \textbf{PPV} & \textbf{NPV} & \textbf{FDR} & \textbf{PSS} & \textbf{ACC} & $\mathbf{\beta=1}$ & $\mathbf{\beta=1.5}$ & $\mathbf{\beta=2}$ & \textbf{MCC} & \textbf{B} & \textbf{CSI} & $\mathbf{F^{<3>}}$ & $\mathbf{F^{<4>}}$ & \textbf{$\kappa$} \\
      \hline
      \textbf{Magnitude} & $\cellcolor{red!25} \times$ & $\cellcolor{red!25} \times$ & $\cellcolor{red!25} \times$ & $\cellcolor{red!25} \times$ & $\cellcolor{red!25} \times$ & $\cellcolor{red!25} \times$ & $\cellcolor{red!25} \times$ & $\cellcolor{green!25} \checkmark$ & $\cellcolor{green!25} \checkmark$ & $\cellcolor{orange!25} \sim$ & $\cellcolor{orange!25} \sim$ & $\cellcolor{orange!25} \sim$ & $\cellcolor{green!25} \checkmark$ & $\cellcolor{green!25} \checkmark$ & $\cellcolor{orange!25} \sim$ & $\cellcolor{orange!25} \sim$ & $\cellcolor{green!25} \checkmark$ & $\cellcolor{green!25} \checkmark$ \\
      \hline
      \textbf{Translation} & $\cellcolor{green!25} \checkmark$ & $\cellcolor{green!25} \checkmark$ & $\cellcolor{green!25} \checkmark$ & $\cellcolor{green!25} \checkmark$ & $\cellcolor{green!25} \checkmark$ & $\cellcolor{green!25} \checkmark$ & $\cellcolor{green!25} \checkmark$ & $\cellcolor{green!25} \checkmark$ & $\cellcolor{green!25} \checkmark$ & $\cellcolor{green!25} \checkmark$ & $\cellcolor{green!25} \checkmark$ & $\cellcolor{green!25} \checkmark$ & $\cellcolor{green!25} \checkmark$ & $\cellcolor{green!25} \checkmark$ & $\cellcolor{green!25} \checkmark$ & $\cellcolor{green!25} \checkmark$ & $\cellcolor{green!25} \checkmark$ & $\cellcolor{green!25} \checkmark$ \\
      \hline
      \textbf{Rotation} & $\cellcolor{green!25} \checkmark$ & $\cellcolor{green!25} \checkmark$ & $\cellcolor{green!25} \checkmark$ & $\cellcolor{green!25} \checkmark$ & $\cellcolor{green!25} \checkmark$ & $\cellcolor{green!25} \checkmark$ & $\cellcolor{green!25} \checkmark$ & $\cellcolor{green!25} \checkmark$ & $\cellcolor{green!25} \checkmark$ & $\cellcolor{green!25} \checkmark$ & $\cellcolor{green!25} \checkmark$ & $\cellcolor{green!25} \checkmark$ & $\cellcolor{green!25} \checkmark$ & $\cellcolor{red!25} \times$ & $\cellcolor{green!25} \checkmark$ & $\cellcolor{green!25} \checkmark$ & $\cellcolor{green!25} \checkmark$ & $\cellcolor{green!25} \checkmark$ \\
      \hline
      \textbf{Noise} & $\cellcolor{red!25} \times$ & $\cellcolor{green!25} \checkmark$ & $\cellcolor{red!25} \times$ & $\cellcolor{green!25} \checkmark$ & $\cellcolor{red!25} \times$ & $\cellcolor{green!25} \checkmark$ & $\cellcolor{red!25} \times$ & $\cellcolor{green!25} \checkmark$ & $\cellcolor{green!25} \checkmark$ & $\cellcolor{green!25} \checkmark$ & $\cellcolor{green!25} \checkmark$ & $\cellcolor{green!25} \checkmark$ & $\cellcolor{green!25} \checkmark$ & $\cellcolor{red!25} \times$ & $\cellcolor{green!25} \checkmark$ & $\cellcolor{green!25} \checkmark$ & $\cellcolor{green!25} \checkmark$ & $\cellcolor{green!25} \checkmark$ \\
      \hline
      \textbf{Time} & $\cellcolor{green!25} \checkmark$ & $\cellcolor{green!25} \checkmark$ & $\cellcolor{green!25} \checkmark$ & $\cellcolor{green!25} \checkmark$ & $\cellcolor{green!25} \checkmark$ & $\cellcolor{green!25} \checkmark$ & $\cellcolor{green!25} \checkmark$ & $\cellcolor{green!25} \checkmark$ & $\cellcolor{green!25} \checkmark$ & $\cellcolor{green!25} \checkmark$ & $\cellcolor{green!25} \checkmark$ & $\cellcolor{green!25} \checkmark$ & $\cellcolor{green!25} \checkmark$ & $\cellcolor{green!25} \checkmark$ & $\cellcolor{green!25} \checkmark$ & $\cellcolor{green!25} \checkmark$ & $\cellcolor{green!25} \checkmark$ & $\cellcolor{green!25} \checkmark$ \\
      \hline
    \end{tabular}}
    \label{table:recapmag1}
\end{table}

\begin{table}
  \centering
  
  \caption{Summary of the metaverification process for the geometric comparison of flood maps. $\times$ are rejected by the criterion, $\checkmark$ are accepted by the criterion, and $\sim$ are in between.}
    \resizebox{0.8\linewidth}{!}{\begin{tabular}{|c|c|c|c|c|c|c|c|c|c|c|c|c|}
      \hline
     & \textbf{H} & $\mathbf{d_{MH}}$ & $\mathbf{P^{2}}$ & $\mathbf{W_{2}}$ & $\mathbf{d_{e}}$ & \textbf{NMI} & \textbf{FSS1} & \textbf{FFS3} & \textbf{FSS5} & \textbf{FSS7}  \\
      \hline
      \textbf{Magnitude} & $\cellcolor{green!25} \checkmark$& $\cellcolor{green!25} \checkmark$ &  $\cellcolor{orange!25} \sim$ & $\cellcolor{green!25} \checkmark$ & $\cellcolor{green!25} \checkmark$ & $\cellcolor{green!25} \checkmark$ & $\cellcolor{orange!25} \sim$ & $\cellcolor{orange!25} \sim$ & $\cellcolor{orange!25} \sim$ & $\cellcolor{orange!25} \sim$ \\
      \hline
      \textbf{Translation} & $\cellcolor{red!25} \times$ & $\cellcolor{green!25} \checkmark$ &  $\cellcolor{red!25} \times$ & $\cellcolor{red!25} \times$ & $\cellcolor{green!25} \checkmark$ & $\cellcolor{green!25} \checkmark$ & $\cellcolor{green!25} \checkmark$ & $\cellcolor{green!25} \checkmark$ & $\cellcolor{green!25} \checkmark$ & $\cellcolor{green!25} \checkmark$ \\
      \hline
      \textbf{Rotation} & $\cellcolor{green!25} \checkmark$ & $\cellcolor{green!25} \checkmark$ & $\cellcolor{green!25} \checkmark$ & $\cellcolor{green!25} \checkmark$ & $\cellcolor{green!25} \checkmark$ & $\cellcolor{green!25} \checkmark$ & $\cellcolor{green!25} \checkmark$ & $\cellcolor{green!25} \checkmark$ & $\cellcolor{green!25} \checkmark$ & $\cellcolor{green!25} \checkmark$ \\
      \hline
      \textbf{Noise} & $\cellcolor{red!25} \times$ & $\cellcolor{green!25} \checkmark$ & $\cellcolor{green!25} \checkmark$ & $\cellcolor{green!25} \checkmark$ & $\cellcolor{green!25} \checkmark$ & $\cellcolor{green!25} \checkmark$ & $\cellcolor{green!25} \checkmark$ & $\cellcolor{green!25} \checkmark$ & $\cellcolor{green!25} \checkmark$ & $\cellcolor{green!25} \checkmark$  \\
      \hline
      \textbf{Time} & $\cellcolor{green!25} \checkmark$ & $\cellcolor{green!25} \checkmark$  & $\cellcolor{orange!25} \sim$ & $\cellcolor{green!25} \checkmark$ & $\cellcolor{green!25} \checkmark$ & $\cellcolor{green!25} \checkmark$ & $\cellcolor{red!25} \times$ & $\cellcolor{red!25} \times$ & $\cellcolor{red!25} \times$ & $\cellcolor{red!25} \times$ \\
      \hline
    \end{tabular}}
    \label{table:recapmag2}
\end{table}

\subsubsection{Correlation analysis}

The number of performance measures to evaluate can be reduced with a correlation analysis to establish relationships between the performance measures and group them. The correlations are identified with graphical representations, such as scatter plots, or Principal Component Analysis (PCA) correlation circles. PCA is a dimensionality reduction technique that transforms the original variables into a new set of orthogonal (uncorrelated) variables called principal components, which capture the maximum variance in the data. The variations of the performance measures for the first two principal components can be visualized on a correlation circle. In this visualization, the lengths of the vectors represent the correlation between variables and the principal components. Correlations between performance measures are identified more visually than in scatterplots by looking at the angle between each vector representing the correlation between the variables.

The scatter plots in Appendices \ref{fig:binary} and \ref{fig:corrshapes} provide a comprehensive view of the pairwise relationships of the performance measures. 
A simpler way to visualize the correlations between the performance measures is to use the correlation circle in Figure \ref{fig:PCA} on the two first dimensions of the PCA (they explain 95\% of the variance).

\begin{figure}[htbp]
    \centering
   \includegraphics[width=0.51\textwidth]{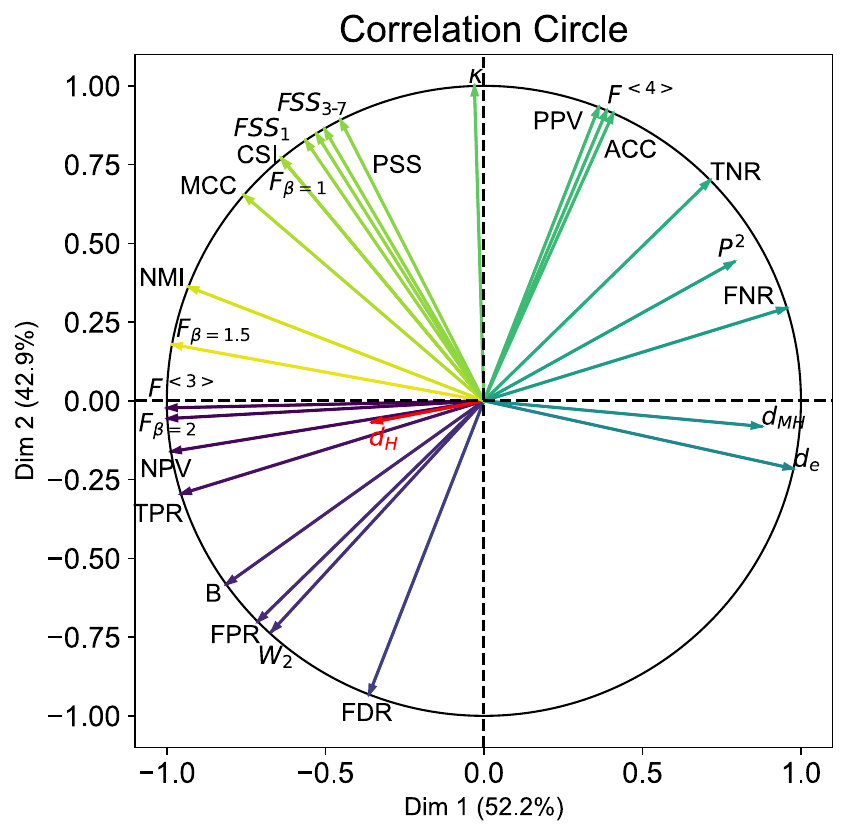}
    \caption{Correlation circle for the two first components of Principal Component Analysis on the performance measures calculated for the reference observation against 4,000 simulations.}
    \label{fig:PCA}
\end{figure}

Contributions to the dimensions differ depending on the performance measures. Except for $d_{H}$ (red arrow in the Figure), all variables show a high correlation with the first two principal components. Linear correlations on the first two components can be identified by interpreting the angle between the arrows. To simplify visualizations, closely correlated variables have low variation of colors. For example, $NPV$, $F_{\beta=2}$, $F^{<3>}$ are closely correlated as well as $PPV$, $F^{<4>}$, and $ACC$.
To validate the linear relationships among performance measures identified on the correlation circle, Figure \ref{fig:pairplot} reports an example of the pairwise relationships between $ACC$ and $F^{<4>}$ which seem correlated and between $ACC$ and $NMI$ which are not.

\begin{figure}[htbp]
    \centering
    \includegraphics[width=0.85\textwidth]{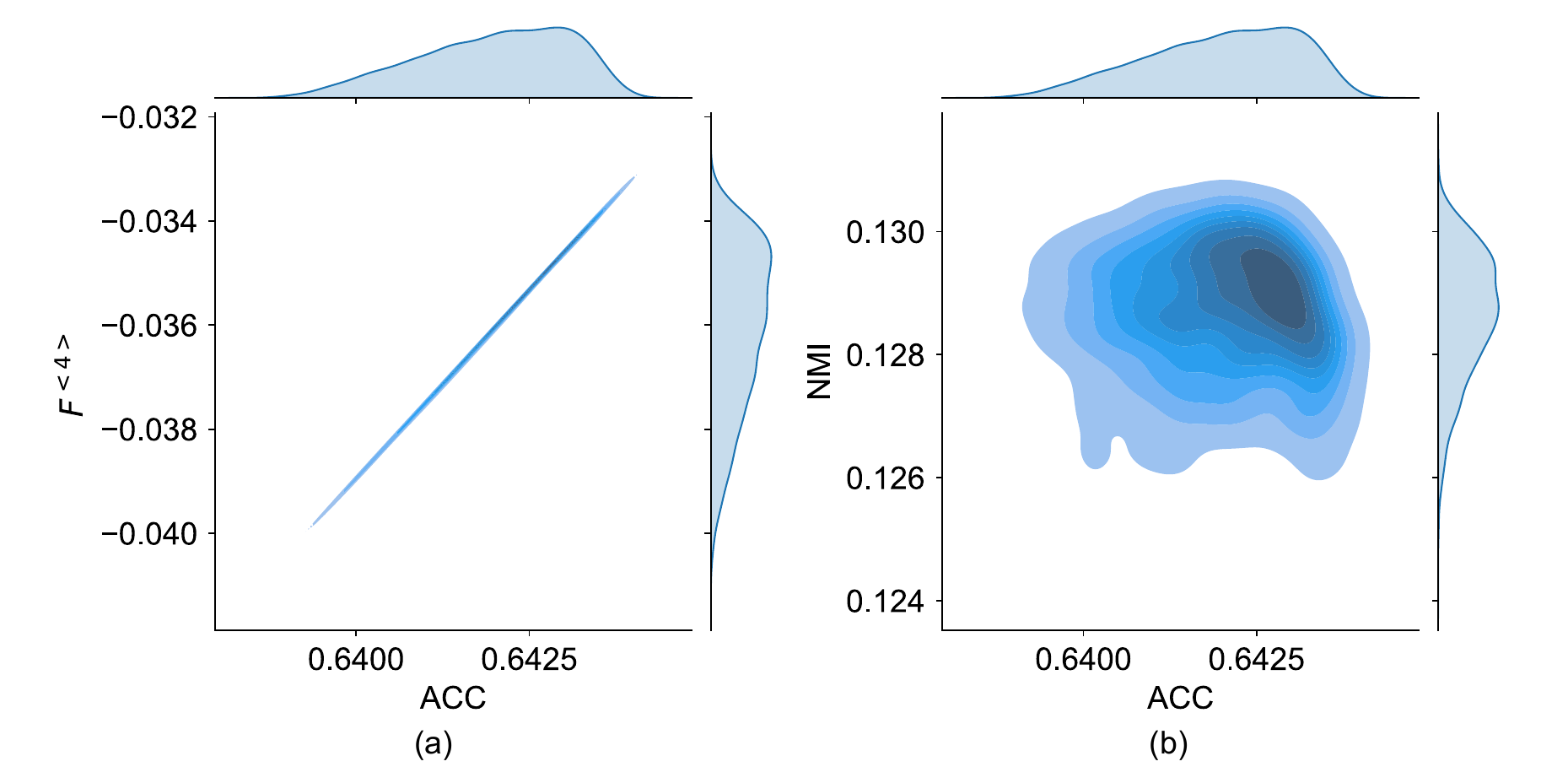}
    \caption{Pairwise comparison of (a) $ACC$ and $F^{<4>}$, and (b) $ACC$ and $NMI$.}
    \label{fig:pairplot}
\end{figure}

The linear correlation is observed between $ACC$ and $F^{<4>}$, but is not between $ACC$ and $NMI$ confirming the analysis of the correlation circle. The main advantage of linear correlations is that a performance measure can be deduced directly from another one, hence reducing the number of performance measures to evaluate. 
Beyond the correlation circle, scatterplots are interesting as they help to identify non-linear relations between the variables, for instance $ACC$ and $F^{<4>}$ are not perfectly linear and contain non-linear relations. The non-linear relations are not exploited in this work.

\subsubsection{Ranking the performance measures}
\label{sec:ranking}

After the selection phase, the practitioner may still have several performance measures available. Then, the last step aims to rank the performance measures. The proposed ranking process is based on the sensitivity of parameter identification to observation errors. 100 flood maps are generated (not more for computational reasons) following the same process as in Section \ref{sec:observed} with uniformly distributed thresholds $(\sigma^{0}_ {j})_{1 \leq j \leq 100}$ between $\sigma^{0}_{1}=-20 \, dB$ and $\sigma^{0}_{100}=-26 \,dB$.  The choice of values is informed based on a proposition of Hostache et al. \cite{hostache2009water} where pixels that are uncertain of being flooded or dry in the yellow and red zones of Figure \ref{fig:Gaussian} are constrained by two thresholds. The two thresholds are: $T_{min}$ the minimum value of dry pixels and $T_{max}$ the threshold that aims to detect all flooded areas at the risk of detecting dry areas with similar $\sigma^{0}$ values. $T_{min}$ is determined with the minimum $\sigma^{0}$ on grassland regions of the dry area in floodplains and $T_{max}$ the maximum value in permanent water bodies. 
For each observation $O^{\sigma^{0}_{j}}$ with $1\leq j\leq 100$, the simulated flood map minimizing the performance measure discrepancy is determined as:
\begin{equation}\label{eq:ranking}
    E_{j}=\min_{1 \leq m \leq M}|\mathcal{L}(O^{\sigma^{0}_ {j}},S(\omega_{m}))-\mathcal{L}(O^{\sigma^{0}_ {j}},O^{\sigma^{0}_ {j}})|.
\end{equation}
Let $E=\{E_{j} | 1 \leq j \leq 100 \}$ describes the set of simulations determined in Equation \ref{eq:ranking}. For each performance measure, the maximum number of times a simulation is identified in the set $E$ is computed as follows:
\begin{equation}
    n_{max}=\max_{1 \leq j \leq 100}| \{E_{j'}|E_{j'}=E_{j}\}|
\end{equation}
where $\{E_{j'}|E_{j'}=E_{j}\}$ denotes the subset of simulations that are equal to $E_{j}$.
The performance measures are ranked according to the $n_{max}$ values. The most robust to observation errors are considered to be the performance measures in which the same simulation is identified most times in $E$.

This ranking process is tested on the 8 remaining performance measures after the selection phase based on the criteria defined in Section \ref{sec:criteria}. For all performance measures, the $n_{max}$ values are calculated and reported in Table \ref{table:nmax}.
\begin{table}
  \centering
  
  \caption{Maximum number of times the same simulation is identified as the closest to the observations (over the 100 observed flood maps) depending on the performance measure.}
   \begin{tabular}{|c|c|c|c|c|c|c|c|c|c|}
      \hline
    & $\mathbf{F^{<4>}}$ & \textbf{ACC} & $\mathbf{d_{MH}}$ & $\mathbf{d_{e}}$ & $\mathbf{\kappa}$ & \textbf{PSS} & \textbf{MCC} & \textbf{NMI} \\
      \hline
    $  \mathbf{n_{max}} $& 76 & 54 & 100 & 54 & 96 & 60& 95& 100 \\
      \hline
      
    \end{tabular}
    \label{table:nmax}
\end{table}
A ranking appears with differences of $n_{max}$ values. $\kappa$, $MCC$, $NMI$ and $d_{MH}$ appear to be the most robust to errors in flood extent identification. Then, for uncertain observations, they seem more adequate. 
For instance, for $NMI$ the same simulated map is optimal for the 100 uncertain observed flood maps. For two of the uncertain flood maps the optimal simulated map for this performance measure is compared with the observations in Figure \ref{fig:compranking}.

\begin{figure}[htbp]
    \centering
    \includegraphics[width=0.85\linewidth]{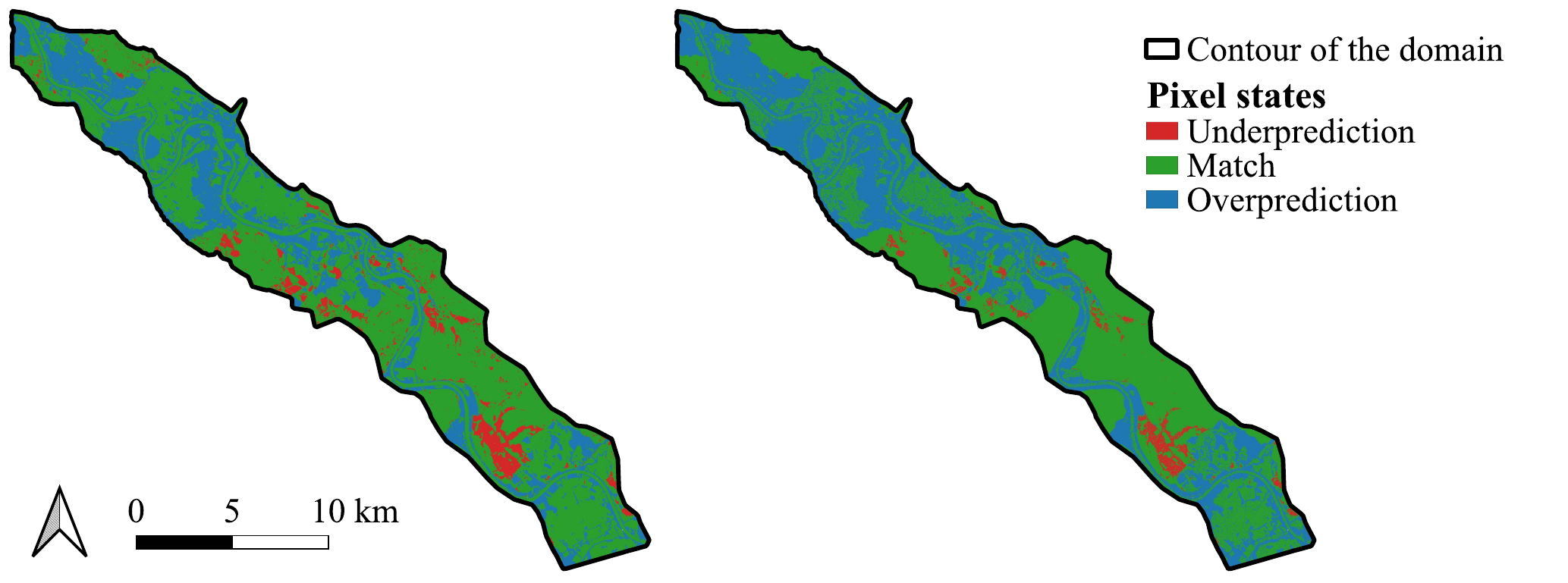}
    \caption{Comparison of the best simulation for $NMI$ with observed flood maps extracted with two thresholds: $\sigma^{0}_{ref}=-20\, dB$ (left), and $\sigma^{0}_{ref}=-26\, dB$ (right).}
    \label{fig:compranking}
\end{figure} 

Although the other performance measures in Table \ref{table:nmax} have lower $n_{max}$ values, it could be interesting to evaluate them, as they are not all linearly correlated with $NMI$. For instance, among the 100 observed flood maps the simulation identified the most time close to the observations differ between $NMI$ and $F^{<4>}$ (variables not linearly correlated). The corresponding simulated maps are compared in Figure \ref{fig:betweenmeasures}.
\begin{figure}[htbp]
    \centering
    \includegraphics[width=0.85\linewidth]{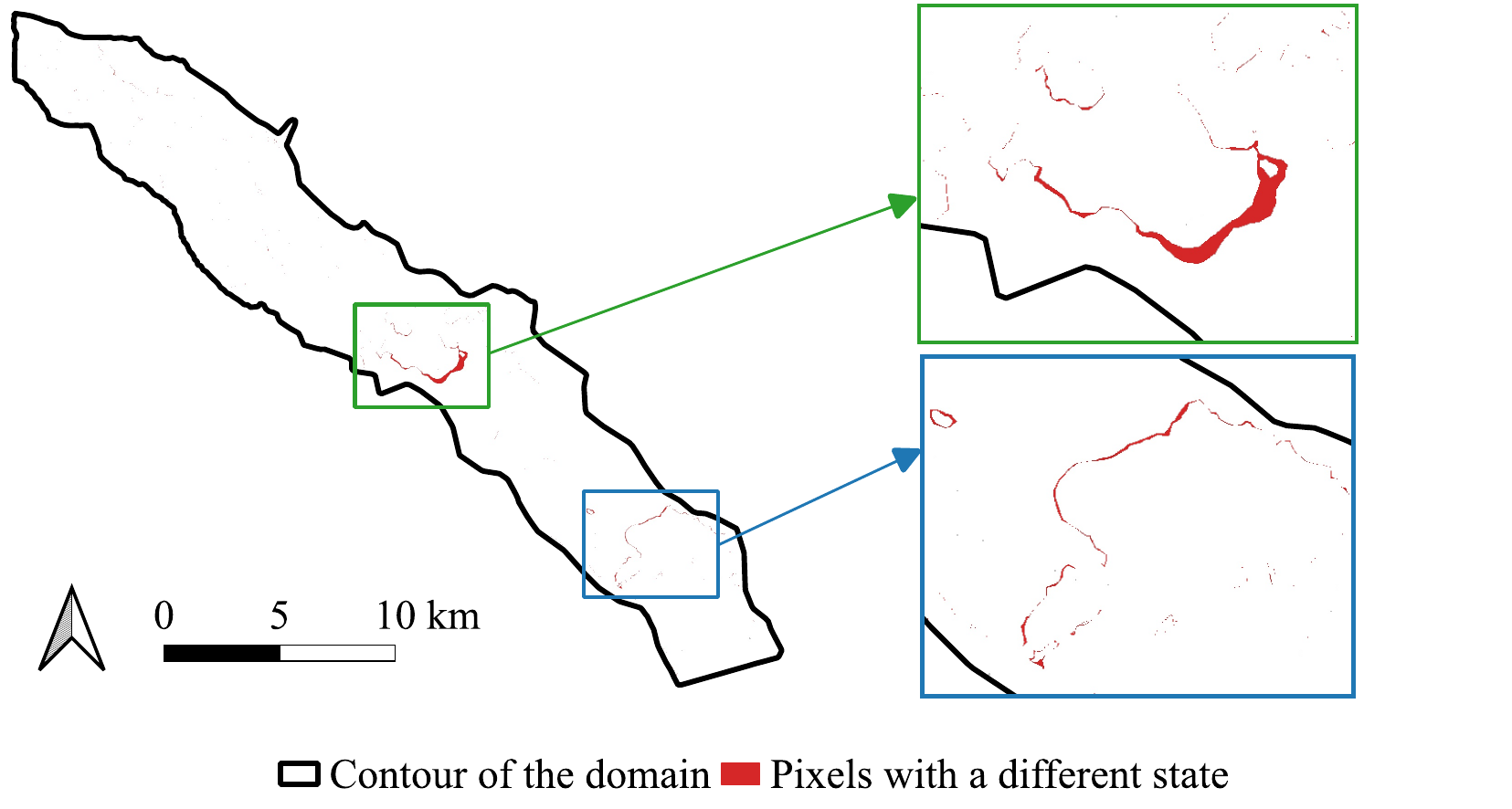}
    \caption{Pixels with different states between the best simulation for $F^{<4>}$ and for $NMI$.}
    \label{fig:betweenmeasures}
\end{figure}
The differences in the Figure are subtle, but the best simulation for $NMI$ has a larger simulated flood upstream than for $F^{<4>}$ which penalizes over-prediction of flooded areas. Then, selecting performance measures may be of interest to have a better representation of the flood in certain areas.

\subsection{Calibration example}
As an illustration, the friction parameter in one of the subdomains is calibrated using one of the performance measures that was highly ranked, namely $NMI$. In Figure \ref{fig:calib}, the distribution of the friction parameter for the 10 \% of the best simulations with respect to the $NMI$ performance measure value is plotted. The comparison of the simulations with the observation with $NMI$ indicates that the friction parameter is more likely in the upper zone of the prior distribution. Then, using this performance measure, it is possible to infer the distribution of the friction in this zone with automatic data assimilation strategies to calibrate the model based on the observations.

\begin{figure}
    \centering
    \includegraphics[width=0.765\textwidth]{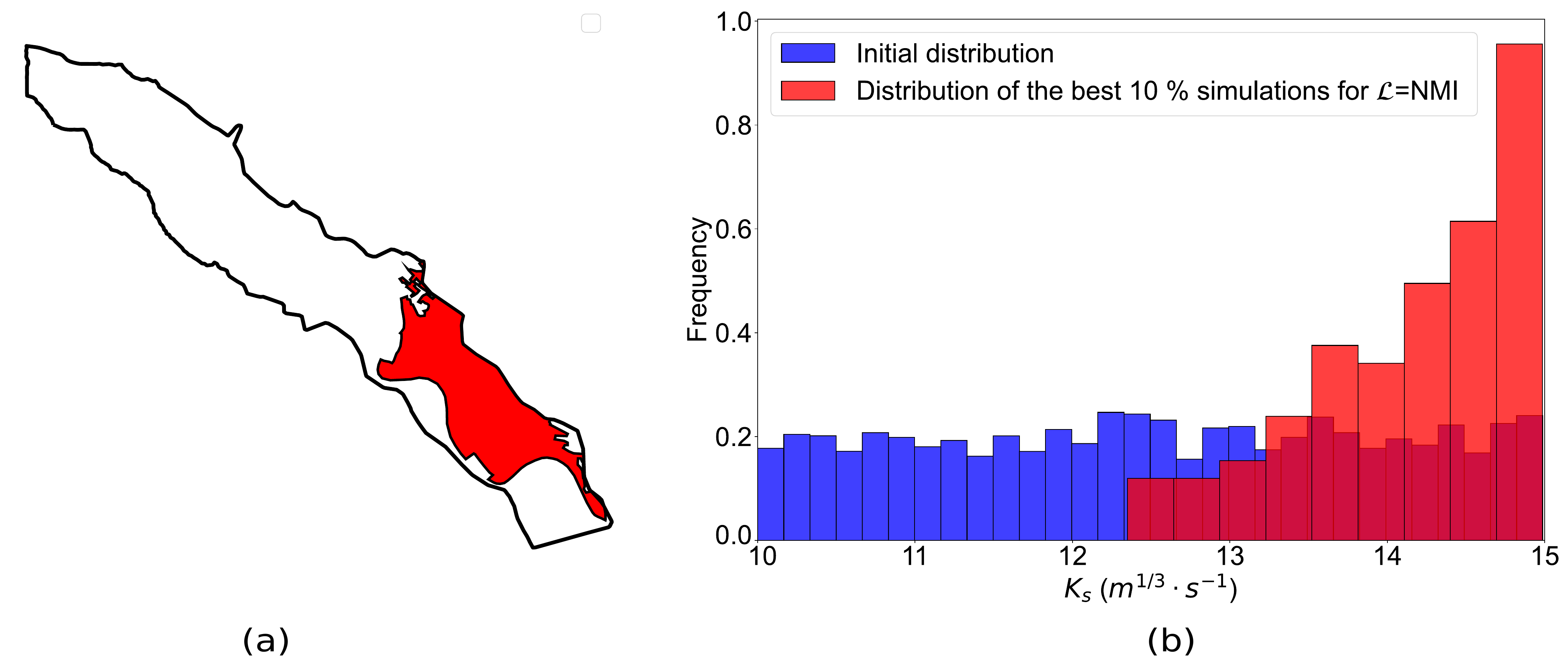}
    \caption{Calibration of one of the zone using $\mathcal{L}=NMI$ performance measure with (a) the zone under study, and (b) the comparison of the initial distribution of $K_{s}$ with the posterior distribution of $K_{s}$ for the best performing simulations.}
    \label{fig:calib}
\end{figure}

\subsection{Limitations}
The criteria from the selection process were proposed based on a literature review, hydraulic processes, and satellite observation characteristics, but may be incomplete to represent all aspects that qualify a flood model against satellite observations.

This work was carried out with observed flood maps extracted with a single extraction procedure (local thresholding). It could be of interest to study the influence of the extraction algorithm on the choice of the performance measure to find out if any changes should be made in the methodology to consider other extraction algorithms.  Furthermore, the simulated fields were interpolated on the observation grid, leading to representativity errors between the simulated mesh and observed grid scales. An additional test may be needed to ensure the choice of the performance measure depending on these scales and the differences induced by the mapping from a triangular mesh to a rectangular grid. In the sensitivity to noise criterion, the noise was considered uniform over the whole image, but could be adapted since the uncertainty of the flood mapping from SAR images is not always spatially uniform with, for example, vegetated or urban areas sometimes mistakenly considered flooded, while these errors are less likely in rural areas \cite{zhao2021deriving}. Another possibility to add meaningful criteria is to consider the two other considerations introduced in \cite{murphy1993good} with the judgements of practitioners on the simulations or criteria on the economic/environmental priorities of the decision makers.

Finally, the methodology is fitted to find a performance measure to compare an observation with a simulation on a global scale. In the case of optimization studies, practitioners may be more interested in getting local information to improve the spatial parameterization of their model. It could be interesting to question the choice of performance measure when comparing local scales and not directly the whole image. For instance, $W_{2}$ performance measure was likely rejected because of the random projections on the grid, but it may be more robust with other projections (e.g. along cross sections of the river). Furthermore, for the comparison of the whole image, some performance measures from the literature were not tested. In any case, the objective of this work remains to propose a methodology for the practitioner to select a performance measure; it allows one to add more performance measures than tested here in the selection process in the future.

\section{Conclusion and Perspectives}

Many performance measures exist to compare spatial fields with different specificities and properties. The practitioner needs to select the most reliable performance measures. Thus, this work proposes a methodology to choose and rank performance measures to qualify a flood numerical model against satellite observations, especially in the case of extracted flood maps. 

The methodology was applied on a study case during a flood event. The performance measure selection process highlighted that some performance measures used in the literature meet more criteria than others. In the study case, it was shown that most performance measures based on the confusion matrix are not equally sensitive to flood extent variations for smaller and larger floods. Although the other criteria did not eliminate many performance measures, each test still eliminated two to three performance measures. Moreover, the analysis of the correlations has shown that many performance measures are closely correlated, hence further decreasing the number of performance measures that should be used when they are redundant. 
Finally, our ranking process was able to identify four performance measures ($\kappa$, $MCC$, $NMI$, and $d_{MH}$) robust at identifying the physical friction parameter. 

In the future, the methodology should be tested with more flood events and experienced with other hydrodynamic models and other flood maps extraction procedures. The selection of a reliable performance measure in a data assimilation framework or to verify 2D simulated fields against 2D observations is essential, and we believe that taking time to analyze the choice of performance measure is important. In hydrology, in the event of an increasing number of spatial missions, new data are available to compare with simulations. The selection of a performance measure for observed/simulated 2D water depth fields could benefit from this work.

\section*{Open Research Section}
The openTELEMAC suite software is freely available online (\url{www.opentelemac.org}). For model construction, topographic and land use data are available on IGN website (\url{https://www.ign.fr/}), and flood hydrographs on HydroPortail (\url{https://hydro.eaufrance.fr/}). Satellite observations are available from the ASF Data Search Vertex platform (\url{https://search.asf.alaska.edu/}) with the pre-processing steps applied in this work.

\acknowledgments
This work was carried out as part of EDF R\&D's MOISE-2 research project on river and coastal flood hazard assessment, whose support the authors gratefully acknowledge. This work was supported and funded by the French National Association of Research and Technology (ANRT) and EDF R\&D with the
Industrial Conventions for Training through REsearch (CIFRE grant agreement 2022/0972). The authors would like to gratefully acknowledge the open-source community, especially that of the openTELEMAC. The authors also express special thanks to Sylvain Biancamaria for the fruitful discussions on satellite imagery. 
%

\appendix
\section*{Appendix A. Extraction procedure of the flood maps}
Water and floods can be detected exploiting that open water surfaces have low backscattering values due to specular reflection and high water permittivity \cite{ulaby1982radar}. A possibility to extract a flood map with pixels classified as dry or flooded is to determine a threshold that separates the two classes. 
The simpler approach is to use global thresholding assuming that the dry and flooded pixels distribution backscattering values are Gaussian distributed. The most-known approach for determining the optimum threshold is the Otsu algorithm \cite{otsu1979threshold}, used, for instance in \cite{schumann2009utility,schumann2010near,pulvirenti2013discrimination}. The algorithm maximizes the difference between the intensity values of the two classes while minimizing the variance within each class by iterating through all possible threshold values. It relies on assumptions such as class Gaussian distribution and both the class sizes and variances should be similar. 

However, the bimodality of the backscattering values is not always identifiable on the whole image and the threshold cannot be directly determined. To address the spatial variability of the threshold to detect water, \cite{chini2017hierarchical,martinis2009towards,twele2016sentinel} propose a split-based approach to use a local-based thresholding algorithm. Local thresholding relies on separating the image into sub-images with bimodal histogram containing large portions of pixels belonging to flooded
and dry classes (usually half water and half dry). Then local thresholds are estimated and combined to calculate an overall threshold for the observation leading to better results in the flood extent detection compared to the global thresholding approach. The implementation in this work is based on the description of the algorithm in \cite{chini2017hierarchical}. They use quad-tree decomposition on the satellite images and combine representative tiles with bimodal Gaussian distributions that contain at least 10\% pixels of each class. The Ashman D coefficient, the Bhattacharyya coefficient, and the surface ratio of the two classes are evaluated on each tile to evaluate the bimodality of the histogram, the Gaussianity of the distributions, and the balance of each class.
The Ashman D coefficient is computed as:
\begin{equation}
    AD(h)=\sqrt{2}\frac{|\mu_{1}-\mu_{2}|}{\sqrt{\sigma_{1}^{2}+\sigma_{2}^{2}}}
,\end{equation}
where $\mu_{1}$, $\mu_{2}$, $\sigma_{1}$ and $\sigma_{2}$ are the distribution means and standard deviations, respectively, and $h$ is the histogram of the sub-tiles. The separability of the Gaussian is ensured for high values of $AD$ and, it is considered in this work and in \cite{chini2017hierarchical} a value of at least 2. 
The Battacharyya coefficient evaluating the similarity of two distributions is computed as:
\begin{equation}
    BC(h,h_{f})=\sum_{k}\sqrt{h(y_{k})}\sqrt{h_{f}(y_{k})}
,\end{equation}
where $y$ are the measurements, $k$ are the bins of the two discrete histograms $h$ and $h_{f}$ where $h_{f}$ is the fitted histogram of $h$.
For a good fit of the normal distributions on the measured histograms, $BC$ should be greater than 0.99. 
Finally, the third condition to consider is that the smaller class should represent at least 10\% of the second class. 

\section*{Appendix B. Convergence of the statistical estimators}\label{appendix:MonteCarlo}

Let $Y=\{\mathcal{L}(O,S(\omega_{m})) |  1\leq m \leq M\} \in \mathbb{R}^{M}$ be a finite discrete random variable representing the values taken by $\mathcal{L}$ between an observation and $M$ simulations. The statistical distributions of $Y$ are analyzed through their moments. The first four moments (mean, standard deviation, skewness, and kurtosis) of the random variable are computed for different number of simulations. The moments of order $k$ are defined as:
\begin{equation}
m_{k}=\sum_{m=1}^{M}\mathcal{L}(O,S(\omega_{m}))^{k}p(X=\mathcal{L}(O,S(\omega_{m})))
.\end{equation}
With a sampling-based approach, the estimators are approximated based on the available sample. To assess the robustness and convergence of the estimators, the bootstrap method is used to estimate the confidence intervals of the estimates. The number of bootstrap samples is set to 500 (after verifying that the estimators confidence intervals do not change for more bootstrap samples). Then, the number of samples needed to reach stable statistical estimates is evaluated with a box plot analysis on the median and confidence intervals of the estimators.

For clarity, since the analysis gives similar results for all performance measures, only the $ACC$ performance measure is arbitrarily presented. Figure \ref{fig:moments} reports the evolution of the first fourth moment for the Accurary ($ACC$) performance measure.  

\begin{figure}[htbp]
\centering
\includegraphics[width= 0.85\linewidth]{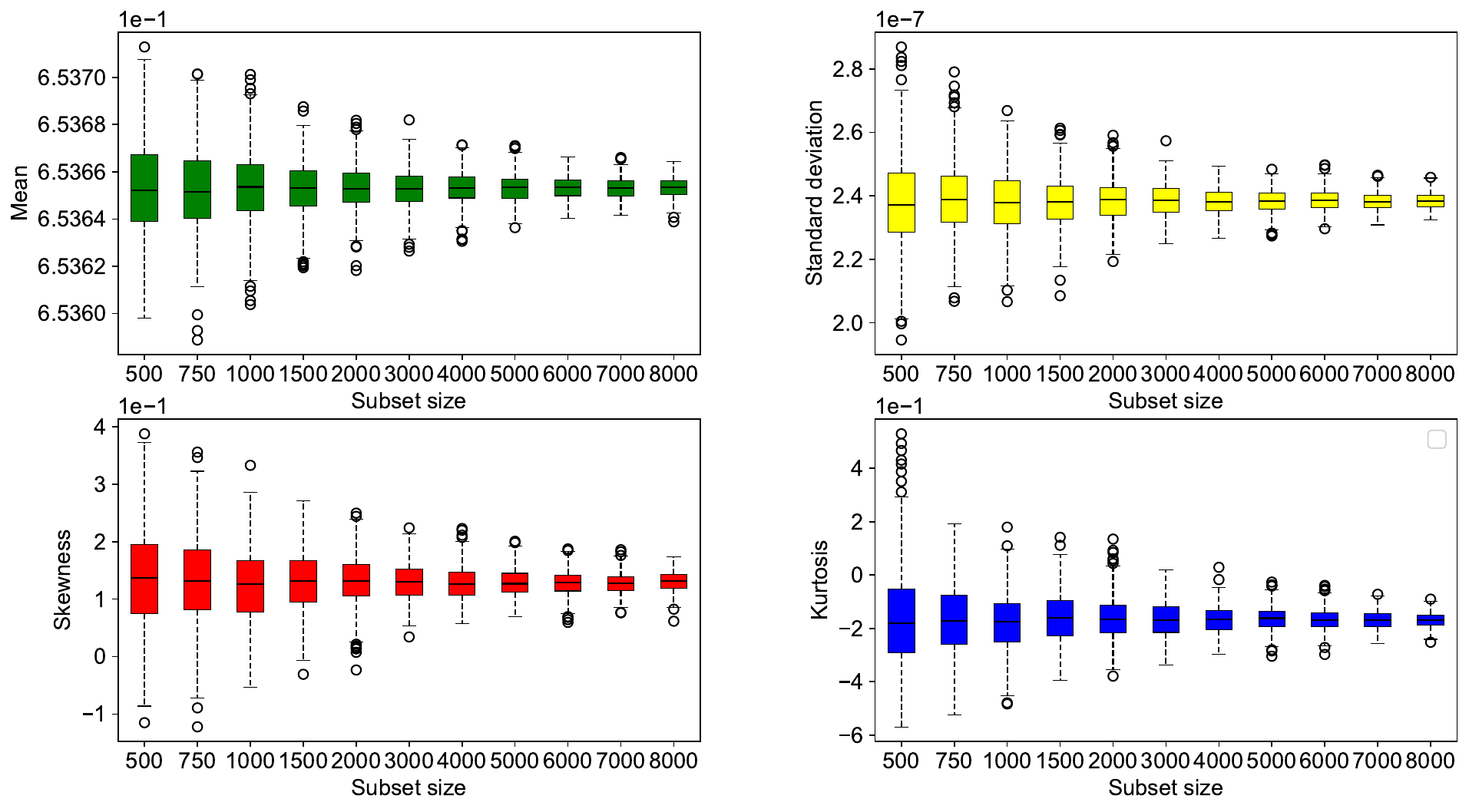}

    \caption{Mean, standard deviation, skewness and kurtosis evolutions for the Accuracy ($ACC$) performance measure on the bootstrap samples.}
    \label{fig:moments}
\end{figure}

\section*{Appendix C. Scatter plots visualization}\label{appendix:additional}
The complete scatter plots comparing all performance measures distributions two by two are visualized in Figures \ref{fig:binary} and \ref{fig:corrshapes}.

\begin{sidewaysfigure}
    \centering
   \includegraphics[width=1\textwidth]{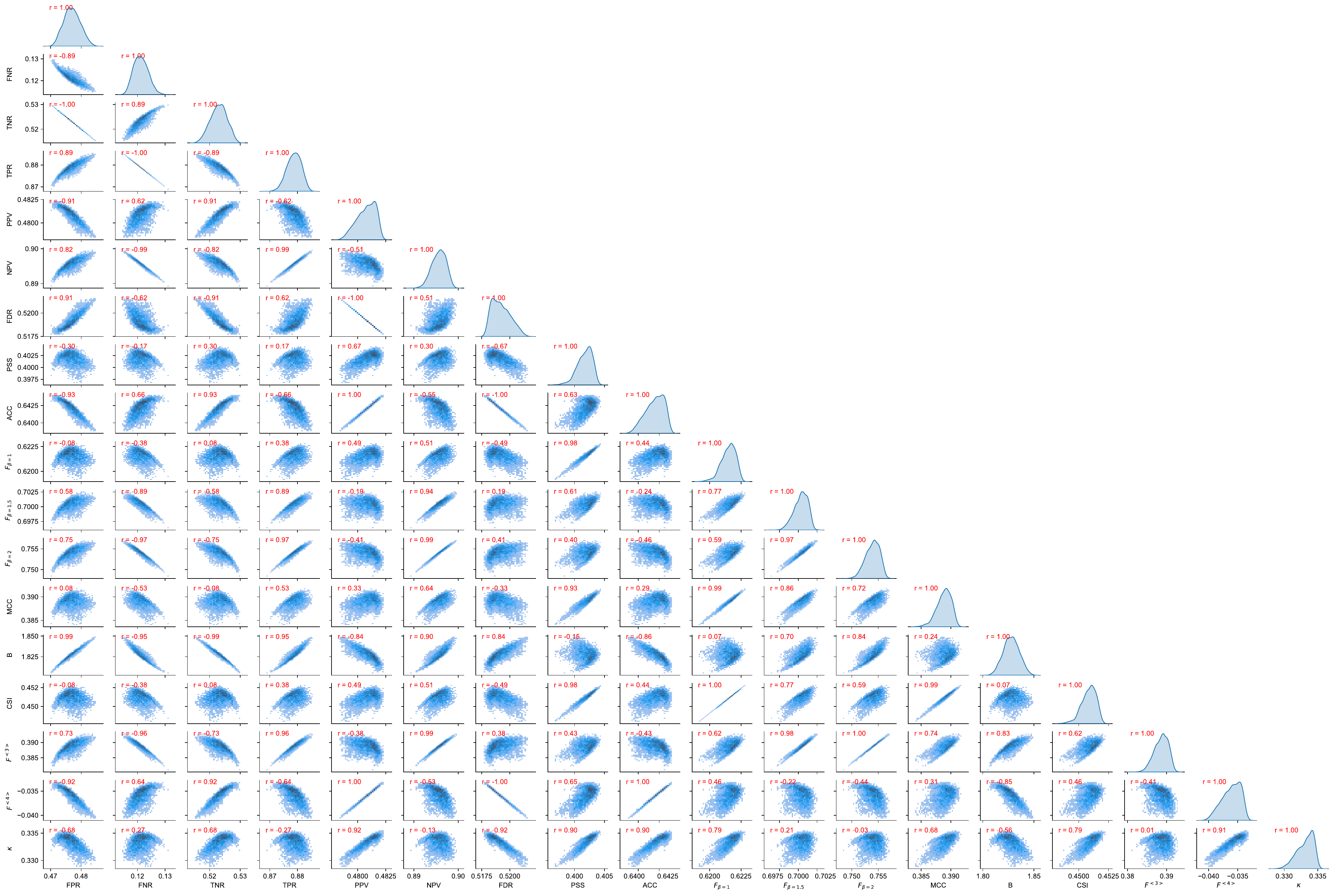}
    \caption{Scatter  plot  for pixel-to-pixel comparison of flood maps.}
    \label{fig:binary}
\end{sidewaysfigure}

\begin{sidewaysfigure}
   
    \centering
   \includegraphics[width=1\textwidth]{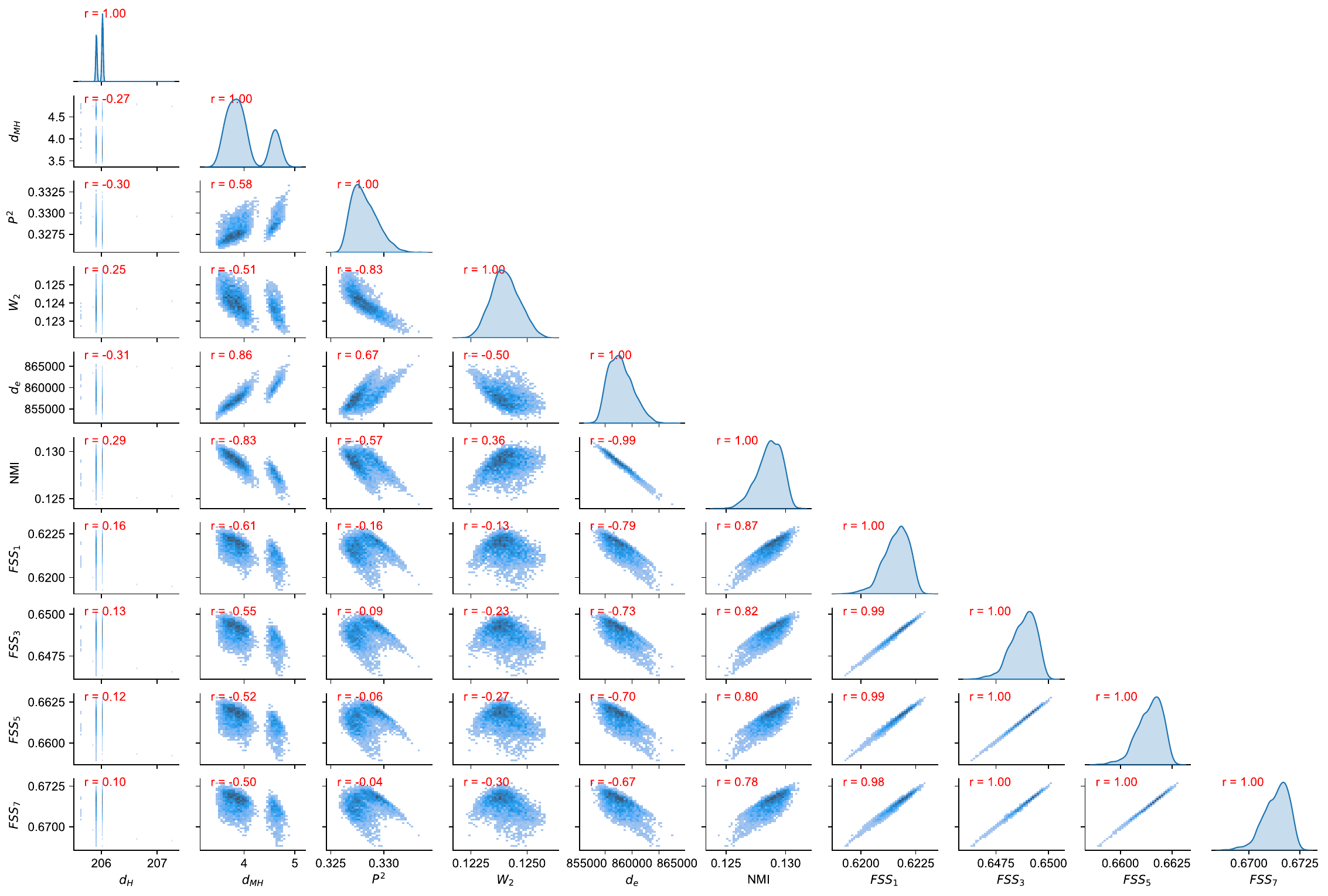}
    \caption{Scatter  plot for geometrical comparison of flood maps.}
    \label{fig:corrshapes}
\end{sidewaysfigure}
%



%
%
\clearpage
\bibliography{template}
\bibliographystyle{unsrt}

%


%
%
%
%
%

\end{justify}
\end{document}